\nonstopmode
\documentclass[9pt,twocolumn,twoside]{osajnl}  %
\journal{josaa}

\usepackage{mathabx}

\setboolean{shortarticle}{false} 
\setboolean{displaycopyright}{true}

\ifthenelse{\boolean{shortarticle}}{\colorlet{color2}{color2b}}{\colorlet{color2}{color2}} 

\title{Large amplitude tip/tilt estimation by geometric diversity for
  multiple-aperture telescopes} 

\author[1]{S. Vievard}
\author[1,*]{F. Cassaing}
\author[1]{L. M. Mugnier}

\affil[1]{Onera -- The French Aerospace Lab, F-92322, Ch\^atillon, France}
\affil[*]{Corresponding author: frederic.cassaing@onera.fr}

\dates{Compiled \today}

\ociscodes{(070.0070) Fourier optics and signal processing, (220.1080) Active or adaptive optics, (010.7350) Wave-front
  sensing, (100.5070) Phase retrieval, (110.5100) Phased-array
  imaging systems, (220.1140) Alignment.}
\doi{\url{http://dx.doi.org/10.1364/ao.XX.XXXXXX}}

\begin{abstract}
A novel method nicknamed ELASTIC is proposed for the alignment of
multiple-aperture telescopes, in particular segmented telescopes. It only needs the acquisition of
two diversity images of an unresolved source, and is based on the computation
of a modified, frequency-shifted, cross-spectrum. It provides a polychromatic
large range tip/tilt estimation with the existing hardware and an inexpensive
noniterative unsupervised algorithm. Its performance is studied and
optimized by means of simulations. They show that with 5000
photo-electrons/sub-aperture/frame and 1024$\times$1024 pixel images, residues
are within the capture range of interferometric phasing algorithms such as
phase diversity. The closed-loop alignment of a 6 sub-aperture mirror provides
an experimental demonstration of the effectiveness of the method. 
\end{abstract}

\newcommand{\W}[1]{\boldsymbol{#1}} 
\newcommand{\V}[1]{\boldsymbol{#1}} 
\newcommand{\VV}[1]{\mathrm{#1}} 
\newcommand{\GP}[1]{\left(#1\right)}
\newcommand{\GC}[1]{\left[#1\right]}

\newcommand{\EST}[1]{\widehat{#1}}
\newcommand{\eqdef}{\stackrel{\triangle}{=}}
\newcommand{\moy}[1]{\left\langle#1\right\rangle}

\newcommand{\NA}{{\mathcal{N}_a}}

\newcommand{\NP}{\mathcal{N}_p} 
\newcommand{\NO}{\mathcal{N}_o} 
\newcommand{\NPH}{\mathcal{N}_{ph}} 
\newcommand{\Vi}{\V{i}} 
\newcommand{\Vj}{\V{j}} 
\newcommand{\Vp}{\V{q}} 
\newcommand{\Vd}{{\V{\delta}}} 
\newcommand{\Wu}{\W{u}} 
\newcommand{\WZ}{\W{Z}} 
\newcommand{\Wp}{\W{p}} 
\newcommand{\Ws}{\W{s}} 
\newcommand{\Wf}{\W{f}} 
\newcommand{\Weps}{\W{\epsilon}} 

\newcommand{\Wx}{\W{x}} 

\newcommand{\Wc}{\W{c}} 
\newcommand{\Wr}{\W{r}} 
\newcommand{\pic}{\mbox{\huge$\curlywedge$}}   

\makeatletter
\newcommand{\sublabel}[1]{\def\@currentlabel{\arabic{section}\Alph{subsection}}\label{#1}}
\makeatother

\begin{document}
\maketitle
\thispagestyle{fancy}

\ifthenelse{\boolean{shortarticle}}{\ifthenelse{\boolean{singlecolumn}}{\abscontentformatted}{\abscontent}}{}

\section{Introduction}\label{sec-introduction}

The resolution of a telescope is ultimately limited by its aperture diameter,
but the size of mirrors is bounded by current technology to about 10\,m on the ground and
to a few meters in space.
To overcome this limitation, interferometry consists in making an array of
sub-apertures interfere; the resulting instrument is called an interferometer
or a multi-aperture telescope.

The sub-apertures can either be telescopes per se, as in current ground-based
interferometers~\cite{VLTI-url,NPOI-url,LBT-url}, or segments of a common
primary mirror, such as in the Keck telescopes, the future extremely large
telescopes~\cite{GMT-url,TMT-url,ELT-url,Colossus-2014}) or large ground collectors
(Cherenkov Telescope Array~\cite{pareschi2013dual}).

\begin{figure}[!b]\centering
		  \includegraphics[width=0.9\linewidth]{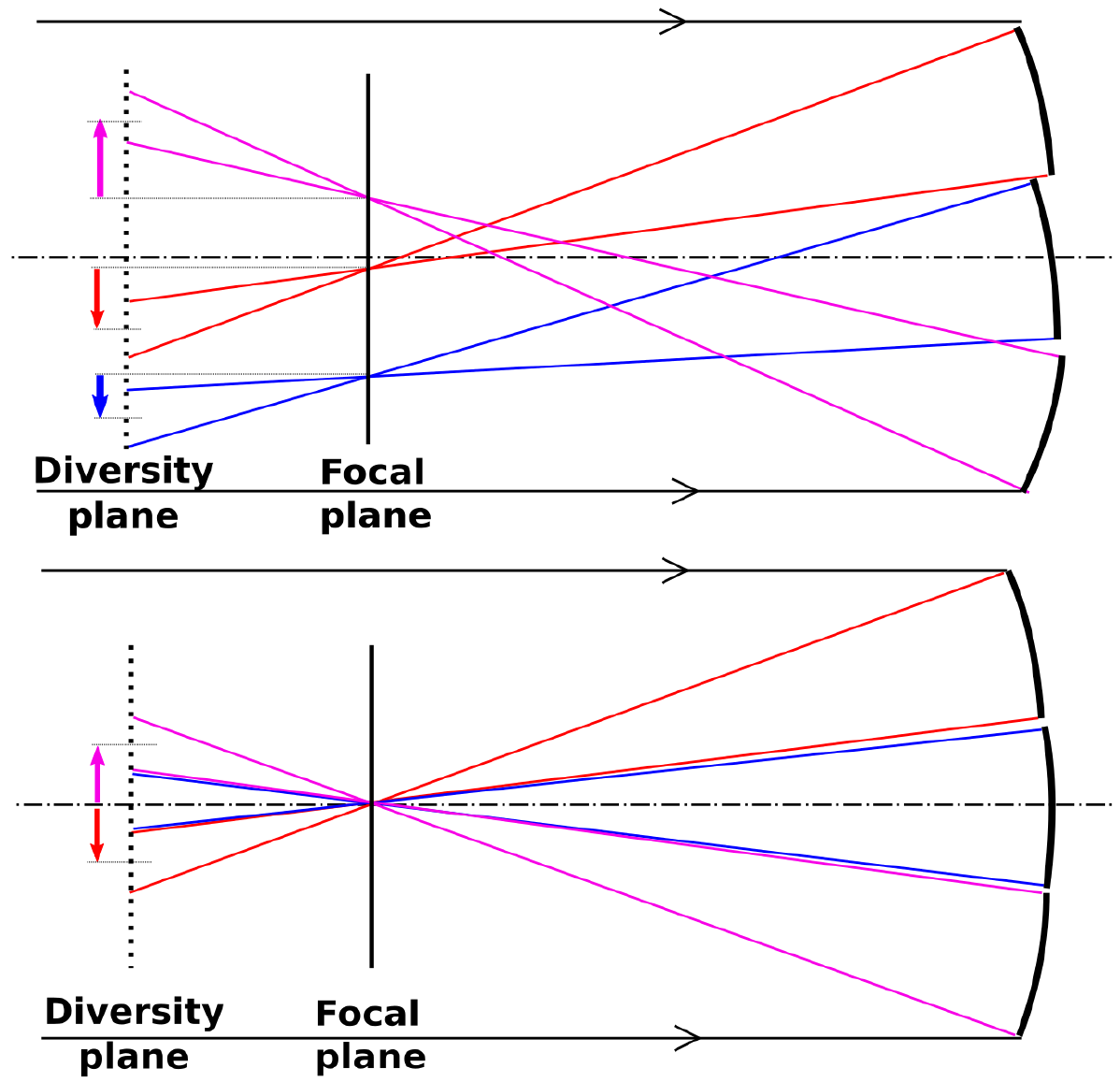}
  \caption{[color on-line] Up: Misaligned telescope: the segment tip/tilts induce
    sub-PSF shifts in the focal plane. An additional shift, specific to each
    sub-aperture, due to the diversity, and denoted by an arrow on the diagram, appears in the defocused plane.
    Down: Aligned telesope. All the sub-PSFs are superimposed in the focal plane. In the defocused plane the additional diversity-induced shift remains.}
  \label{instrum_PD}
\end{figure}

So far, this technique has been operational only on ground-based telescopes,
but interferometers have long been forecast for high-resolution spaceborne
astronomy, for the soon-to-be-launched James Webb Space Telescope (JWST)~\cite{JWST-url} or other
projects~\cite{DARWIN-url,bolcar-LUVOIRtechno-spie15,TALC_sauvage,pitman2004remote},
and for Earth observation~\cite{Mugnier-p-04,Mesrine-p-06}.

To reach the diffraction limit, all the sub-apertures of such a telescope
must be phased to within a small fraction of the wavelength. A critical
sub-system of interferometers is thus the Cophasing Sensor (CS), whose goal is
to measure with this kind of precision the relative positioning errors between
the sub-apertures (differential piston and tip/tilt), which are the specific
low-order aberrations of an interferometer and the main sources of wave-front
degradation (Fig.~\ref{instrum_PD}).

Focal-plane wavefront sensing is an elegant solution to measure the wavefront
degradation. Since the near-focal images of any source taken by
a 2D camera show distortions when the telescope is not aligned
(Fig.~\ref{instrum_PD}), these misalignments can be retrieved by solving the
associated inverse problem.  The phase retrieval technique, based on the analysis of
the sole focal-plane image, is generally not sufficient to retrieve piston and
tip/tilt without ambiguity except in specific cases~\cite{Baron-a-08}.  The
phase diversity
technique~\cite{Gonsalves-PhaseXretrieval_diversity,Mugnier20061}, typically
based on a focal and a slightly defocused images, removes all ambiguities and
operates even on unknown extended sources.

The fact that phase diversity can be used as a CS on a segmented aperture
telescope was recognized very early~\cite{Paxman-88}, and extensively studied
for the JWST~\cite{Redding-p-98,Carrara-p-00,LLee-p-03,Dean-p-06}, for
Darwin~\cite{Mocoeur-p-06b}, and for the European Extremely Large Telescope (E-ELT)~\cite{Meimon-p-08a} in
particular. Additionally, in contrast with most pupil-plane-based devices, phase
diversity enjoys three appealing characteristics: firstly, it is appropriate
for a large number of sub-apertures, because the hardware complexity remains
essentially independent of the number of sub-apertures.  Secondly, the CS is
included in the main imaging detector, simplifying the hardware and minimizing
differential paths. Thirdly, it can be used on very extended objects. These
properties are strong motivations for the choice of phase diversity as a CS,
even when looking at an unresolved source.

The measurement and correction described above of piston-tip-tilt by means of
a CS to within a small fraction of $\lambda$ is hereafter called the \emph{fine
phasing mode}. This mode assumes
that the tip/tilts are smaller than typically $\lambda/8$ (see Eq.\ref{eq-conv-tilt}) and the differential
pistons are within the coherence length (cf~\cite{Baron-a-08}).  However, during the first
alignment steps (after integration or deployment), the disturbances are much
larger than a few micrometers. 

A preliminary \emph{geometrical alignment mode} is thus mandatory, to efficiently
drive the telescope into the reduced capture range of an interferometric
CS. This mode consists in an incoherent superimposition of the focal plane images
from each sub-aperture. Only then can the relative piston errors between
sub-apertures be measured and corrected for the fine phasing to operate.
Using only one image, such as the common focal image of
an unresolved source, the geometrical alignment is impossible at least when all
sub-apertures are identical since their Point Spread Functions (or sub-PSFs) all have the same shape. Even
if these sub-PSFs are well separated in the field with clearly identifiable
positions, it is impossible to associate them with their respective sub-apertures. 

The solution selected by JWST to identify the sub-PSFs is temporal
modulation~\cite{JWST_WFSC_2006}, but the measurement time scales with the
number of sub-apertures.  A method called Geometrical Phase Retrieval (GPR),
with the ability to increase the JWST fine algorithm capture range, was
developed~\cite{Thurman-JOSAA-11}, refined~\cite{JurlingFienup-JOSAA-14},
experimentally demonstrated~\cite{Carlisle-ApplOpt-15} and should be
implemented in the JWST~\cite{JWST_preparing_WFSC}. Based on geometrical
optics, it uses typically four to six defocused images per segment.  The JWST
geometrical alignment mode, called image-stacking
operation~\cite{acton2012wavefront}, was originally forecast to last around 1~week of commissioning time, but
the GPR should provide substantial time savings~\cite{Carlisle-ApplOpt-15}.

In this paper, a novel method for the geometrical alignment of the subapertures is
proposed. The alignment error is encoded by each sub-PSF's shift in the focal plane image. As can be seen on Figure~\ref{instrum_PD}, a defocus induces an additional shift of the sub-PSFs in the diversity plane with respect to the focal plane. Each of these additional shifts is given by the defocus, and is different for each sub-aperture. Using these two images jointly, our method extracts the position and identifies the sub-aperture sub-PSF, thanks to a modified cross-correlation.
Based on geometrical optics similarly to GPR, it thus only needs two images of an
unresolved source and a simple data processing. 
Additionally, its capture range is only limited by the imaging
sensor field of view.  

In Section~\ref{sec-deriv}, a closed-form model of the
multi-aperture Optical Transfer Function (OTF) is derived.
Section~\ref{sec-elastic1} provides a solution for large amplitude
tip/tilt measurement from a modified cross-spectrum of two diversity
images. The method is then validated and optimized by simulation in Section~\ref{sec-optim-perf}.
Finally, in Section~\ref{exp-valid}, an experimental validation is performed on a segmented mirror.

\section{Closed-form model for the multi-aperture OTF  with geometric diversity}
\label{sec-deriv}

Throughout this paper, the considered multiple-aperture telescope is based either
on a segmented telescope or on a telescope array, all feeding a common
focal-plane image detector and operates at a central wavelength $\lambda$ with
a focal length $F$. We assume the object is a point source at infinity. In
this section, the telescope's pupil, its disturbances and its OTF
are successively introduced. 

\subsection{Model of the multiple aperture}

The pupil of the telescope is assumed for simplicity to be made of a set of $\NA$ circular
sub-apertures (of index $n$ ranging from~1 to $\NA$) with identical radius
$R$. In the following we use the reduced coordinate $\Wu$, defined as the
usual coordinate $\Wr$ in the pupil plane divided by $R$ ($\Wu=\Wr/R$).
Similarly we
define $\Wc_n$ as the reduced coordinates of the center of each sub-aperture.
In each of the two so-called diversity planes (of index $d \in \{1,2\}$), each
sub-aperture is characterized by its complex amplitude transmittance
$\Wp_{n,d}$ whose shape is described by the disk function $\Pi$,
\begin{equation}
\Pi(\Wu)=\left\{
\begin{array}{lll}
1  & \mbox{for}\,\, |\Wu| \leq 1, \\
0 & \mbox{elsewhere,}
\end{array}
\right.
\end{equation}
and the pupil transmission is constant over each sub-aperture, equal to
$\rho_n$. The phase of $\Wp_{n,d}$ is the sum of unknown (and seeked) phases
plus known diversity phases $\Psi_{n,d}(\Wu)$, and is expanded on a local
orthonormal basis of Zernike polynomials~\cite{Noll-ZernikeXpoly}. A description of the first $4$~modes is
given in Appendix~\ref{app-zern}. The amplitude, in wave unit, of the
$k$\textsuperscript{th} mode over the $n$\textsuperscript{th} sub-aperture is
defined as $a_{k,n}$. Since the specific aberrations of a multi-aperture
telescope are piston ($k=1$), tip and tilt ($k=2$ and $k=3$), the maximal value of $k$
considered here is $3$. Each sub-aperture transmittance in
the $d$\textsuperscript{th} diversity plane can then be written as:
\begin{align}
\Wp_{n,d}&=\GC{\rho_n\Pi_{n,d}\exp\GP{j\sum_{k=1}^32\pi a_{k,n}\WZ_{k}}}\star\delta_{\Wc_n},
\label{equpuptot}\\
\text{with~} \Pi_{n,d}&=\Pi\,\exp\GC{j\Psi_{n,d}} \text{~and~} j^2=-1.
\end{align}
where $\star\delta_{\Wc_n}$ denotes the shift operator by vector $\Wc_n$.
The total pupil transmitance in
the $d$\textsuperscript{th} diversity plane is then :
\begin{equation}
\Wp_d=\sum_{n=1}^{\NA}\Wp_{n,d}.
\label{pupilletot}
\end{equation}

\begin{figure}[!ht]\centering
  \includegraphics[width=\linewidth]{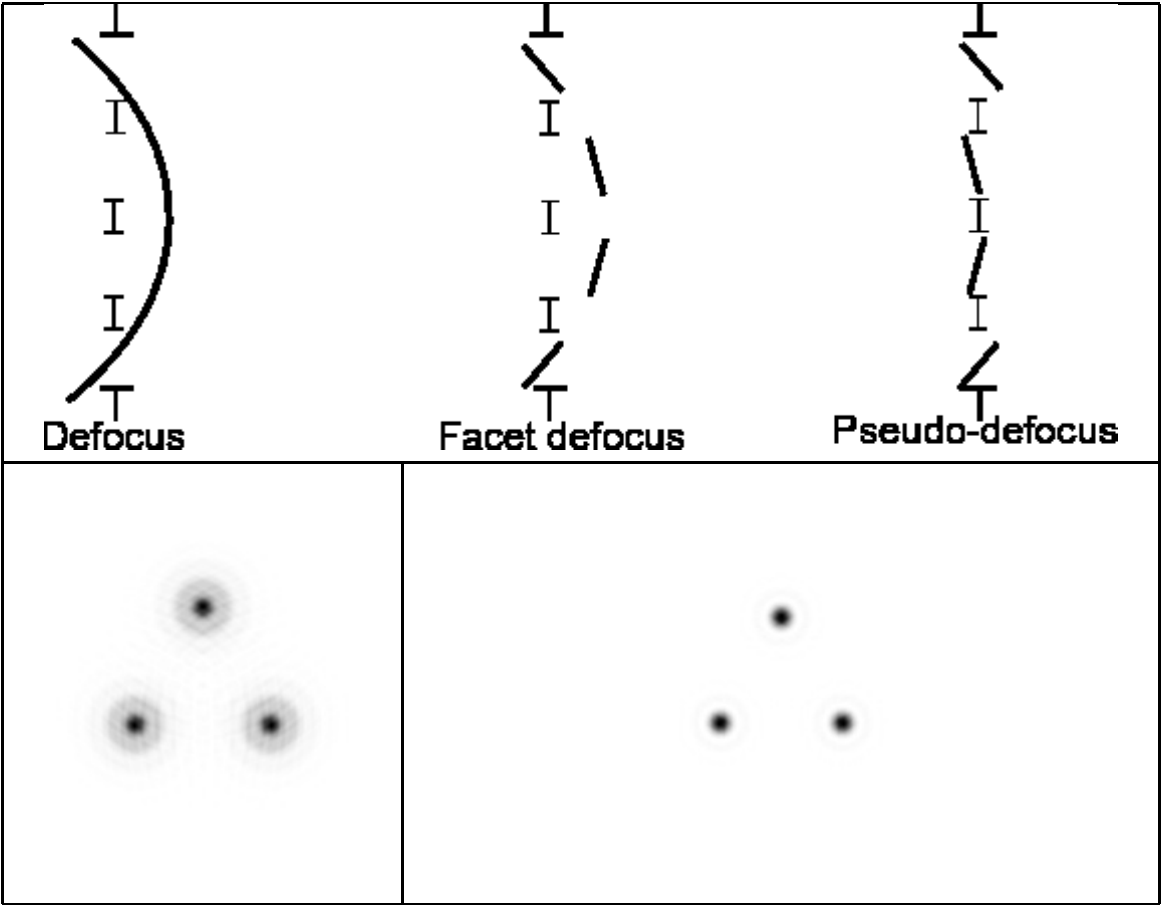}
  \caption{Up: From left to right: global defocus over the full aperture ; facet
    defocus as piston/tip/tilt over each sub-aperture, pseudo-defocus as
    tip/tilt over each sub-aperture.
    Down: Left: Defocused image with a $10~rad$ amplitude. The sub-PSFs are sprawled. Right: Pseudo-defocus or facet defocus  with a $10~rad$ amplitude. The sub-PSFs sprawling is prevented because the facets are plannar.}\label{defoc-pseudodefoc}
\end{figure}

There are several ways
to acquire these diversity images. A first solution is global temporal
modulation: a known full-aperture aberration is applied prior to image
acquisition; a usual example is the use of the focal image (no aberration) and
a defocused image (by a longitudinal shift of the detector or of a mirror, or
by lens
insertion through a filter wheel~\cite{Carlisle-ApplOpt-15}). A second
solution is spatial modulation: the beam is split in two parts, routed to
simultaneously form two distinct images with different aberrations (e.g. focus,
astigmatism) on two different (parts of the) detector(s) \cite{Cassaing-p-06b}.
A third solution is local temporal modulation (Fig.~\ref{defoc-pseudodefoc}):
a global defocus is approximated by its piston/tip/tilt projection on the
sub-apertures, or even to the sole tip/tilt part: the latter
will be called \textit{pseudo-defocus} from now on. This
pseudo-defocus can be introduced simply, either by moving the actuators
intended to correct for the sub-aperture misalignment, or by an array of
(achromatic) prisms in the filter wheel. This provides the sub-PSF's sprawl 
induced by the global defocus, and allows to exclusively acquire focal plane images.

\subsection{Closed-form model of the OTF}\sublabel{deriveOTFPSF}

We define the reduced OTF $\Ws_{d}$ of the $d^{th}$ diversity plane as the autocorrelation ($\otimes$) of
the reduced pupil transmittance $\Wp_d$:
\begin{equation}
\Ws_{d}(\Wu)=(\Wp_{d} \otimes \Wp_{d})(\Wu)=\int
        \Wp_{d}(\Wu')\,\Wp_{d}^*(\Wu+\Wu')\,d\Wu'. \label{autocorr}
\end{equation}
Because the support width is doubled by autocorrelation, $|\Wu|$ ranges
between $0$ and $2$ in the OTF planes. The OTF with the usual spatial frequency $\Wf$ in
$rad^{-1}$ units in the object frequency space is thus $\Ws_d(\Wf\lambda/R)$.
Merging Eqs.~(\ref{equpuptot}) and (\ref{autocorr}), the global
OTF $\Ws_{d}$ in the $d^{th}$ diversity plane is then written as:
\begin{equation}
\Ws_d(\Wu)= \sum_{n=1}^{\NA}\sum_{n'=1}^{\NA}\GC{(\Wp_{n,d}\otimes\Wp_{n',d})
              \star\delta_{(\Wc_n-\Wc_{n'})}}(\Wu).\label{OTF}
\end{equation}

The telescope's PSF can be obtained by computing the Inverse Fourier Transform of
the OTF. Eq.~(\ref{OTF}) shows that the OTF is the sum of $\NA^2$ terms: $\NA$
so-called photometric terms (when $n=n'$, centered at $\Wu=\W0$), and $\NA(\NA-1)$
interferometric terms (when $n \ne n'$, centered at
$\Wu_{n,n'}=(\Wc_n-\Wc_{n'})$). The photometric terms are the autocorrelations
of each sub-aperture, and the interferometric terms are the correlations
of two different sub-apertures.  In this paper dealing with large amplitude
alignment, the interferometric terms will be discarded for 4 reasons. Firstly,
when large tip/tilt errors are present, it is very likely that large piston
errors are also present. When their amplitude is larger than the coherence
length (a few micrometers for broadband observations), interferometric terms
are dimmed by the coherence enveloppe. Secondly, interferometric terms are
also dimmed by significant differential tip/tilts. Thirdly, at this stage of
operation, there is no need to see the high frequency contents of the image,
only the sub-PSF positions are of interest; the detector can thus be rebinned
to reduce the number of pixels to process. And lastly, if all these natural
filters are not sufficient, an explicit low-pass filter can be used to isolate
the superimposed photometric terms (in the center of the Fourier plane) from
the fringe peaks.

An important result is that in presence of tip/tilt errors, the photometric
terms from each sub-aperture are affected by a phase ramp, with the same
slope than their associated pupil transmittance $\Wp_{n,d}$ \cite{Baron-a-08}.
This can be simply understood by the fact that the Fourier Transform (FT) of the two gives the
focal signal (in intensity or amplitude) at the same location. 
The Zernike polynomials were previously used over sub-apertures with a
unit radius. In order to use the same modes (Appendix~\ref{app-zern}), Zernike
polynomials $2$ and $3$ with a doubled support (but same slope) will be written
$2\WZ_{k}\GP{\Wu/2}$. Keeping only incoherent terms, the $d^{th}$ diversity plane OTF writes:

\begin{align}\label{eq-otf-incoh-complete}
\Ws_d(\Wu)= & \sum_{n=1}^{\NA}\rho_n^2\Lambda_{n,d}(\Wu)\exp{\GC{j\sum_{k=2}^{3}4\pi
    a_{k,n}\WZ_{k}\GP{\frac{\Wu}{2}}}}, \\
\text{with~}\Lambda_{n,d}=&\GP{\Pi\exp\GC{j\Psi_{n,d}}}\otimes
                          \GP{\Pi\exp\GC{j\Psi_{n,d}}}.
\end{align}

\begin{figure}[!ht]
\includegraphics[width=\linewidth]{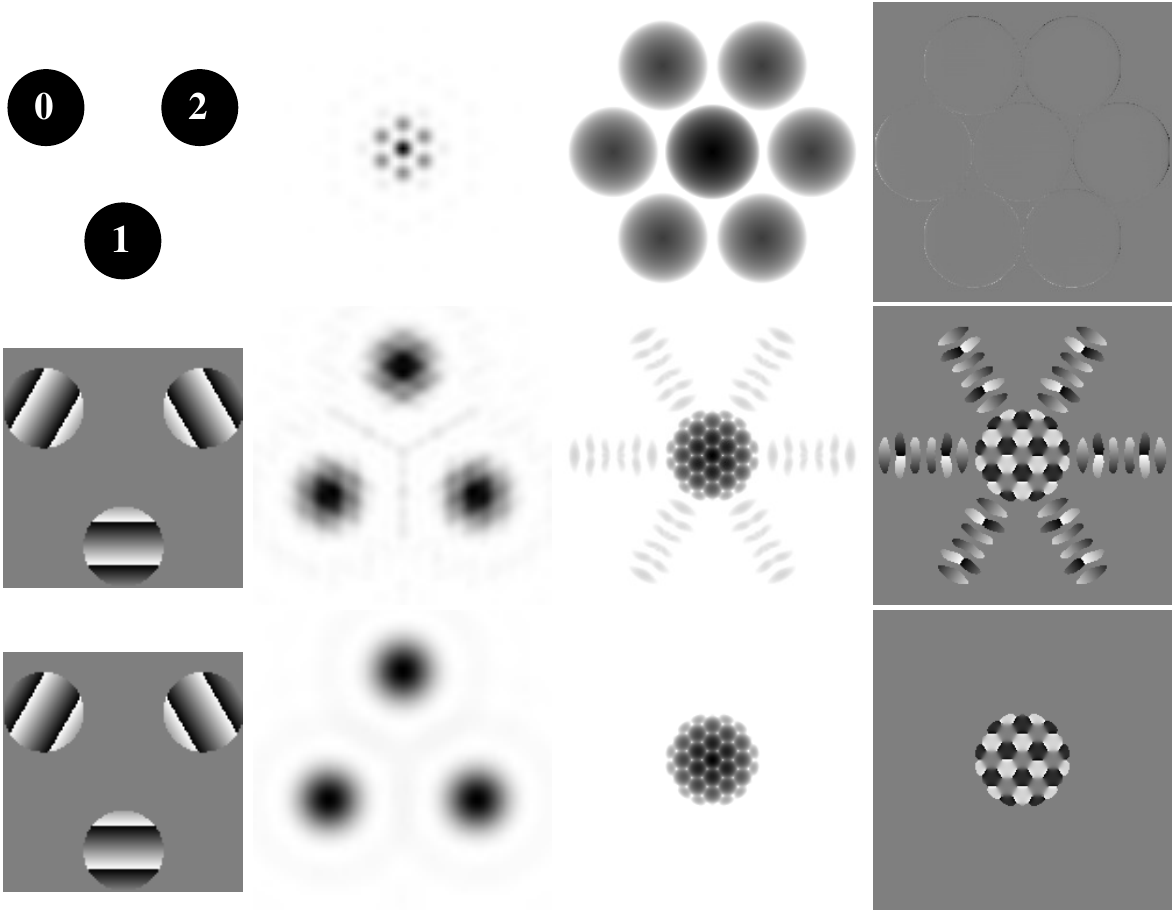}
  \caption{From left to right column: pupil, PSF, OTF modulus (logarithmic scale) and phase for a
    3~sub-aperture configuration (see text and appendix for more details): 1)
    No aberration, coherent sub-PSFs; 2) Tip/Tilt aberrations on pupil 0, 1
    and 2, coherent sub-PSFs;
    3) Tip/Tilt aberrations on pupil 0, 1
    and 2, incoherent sub-PSFs. }\label{fig-illustration}
\end{figure}

To illustrate the previous derivations, Fig.~\ref{fig-illustration} presents
the pupil, the PSF, the OTF modulus and the OTF phase of a 3~sub-aperture
telescope. Line~1 is the no-aberration coherent case. All sub-PSFs are
superimposed in the focal plane, and the OTF modulus is the sum of 9~terms:
6~interferometric terms (two for each baseline) with a chinese-hat peak shape
(hence the symbol $\Lambda=\Pi\otimes\Pi$) and 3~superimposed photometric
terms which form a stronger central peak. In line~2 the impact of
tip/tilt aberrations on the sub-apertures is highlighted. The sub-PSFs are pulled
apart in the focal plane. Although some weak fringes remain
present in the image, the OTF interferometric peaks are strongly attenuated, as previously mentioned. 
Line~3 illustrates the effect of the same previous aberration in an incoherent
case, cancelling all interferometric terms. The information in the
photometric peak of the OTF is not modified with respect to the coherent case. 

In conclusion, the sum of all
superimposed photometric peaks in Eq.~(\ref{eq-otf-incoh-complete}) creates an
intricate pattern (Fig.~\ref{fig-illustration}~lines~2~and~3), with a
different value in each diversity plane, from which the
seeked unknown tip/tilts of each sub-aperture must be estimated. Such an
estimator is derived in the next section.

\section{ELASTIC, a sub-aperture tip-tilt estimator}
\label{sec-elastic1}

Before the quantative mathematical derivation in
subsection~\ref{sec-elastic-direct-model}, the physical origin of this large
amplitude tip-tilt estimator is explained in
subsection~\ref{sec-elastic-principle}.

\subsection{Principle of the estimator}\sublabel{sec-elastic-principle}
Figure~\ref{instrum_PD} shows that the focal image is only sensitive to the
tip/tilt from each sub-aperture, whereas in the diversity image the seeked
tip/tilt is combined with an additional shift resulting from the diversity
(here a defocus obtained by longitudinal shift of the detector). If
the diversity image is telecentric, then this additional shift does not depend
on the sub-aperture tip/tilt, but only on the introduced diversity. The same
holds for a temporal diversity induced by the
sub-aperture actuactors, such as a pseudo-defocus.
If these diversity-induced shifts over all the sub-apertures are sufficiently
different, then this differential information between the two diversity images
enables the association of the sub-PSFs to their sub-aperture index. The goal is now to
convert this qualitative visual process to an unsupervised quantitative
measurement, using a simple algorithm that can be operated whatever the number
of sub-apertures is or their relative positions are.

\begin{figure}[!hb]\centering
  \includegraphics[width=\linewidth]{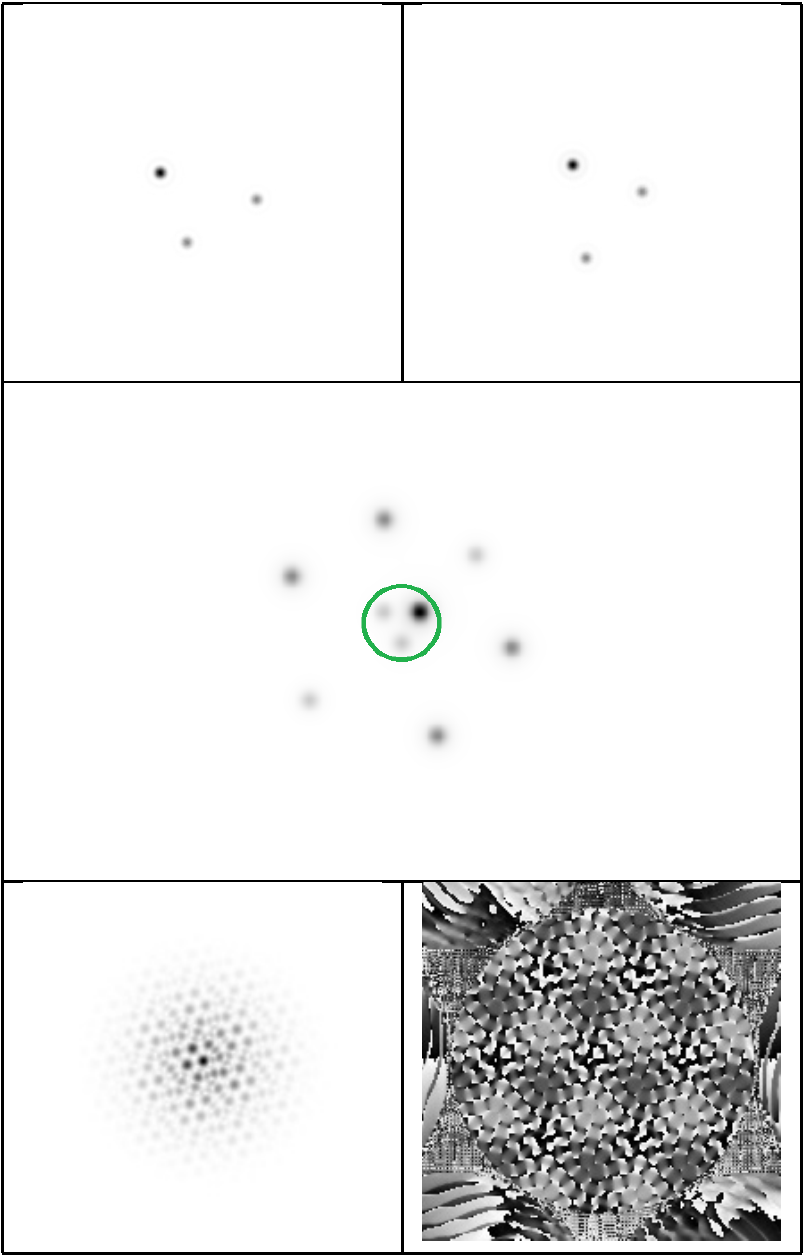}
  \caption{Illustration of ELASTIC. Line~1: two diversity images with a
    3-aperture interferometer (left: focal
    plane with tip/tilt errors, right: with an added pseudo-defocus). Line~2: Image
    correlation. The 3 autospots are inside the overplotted circle in the center while the 6 interspots are
    scattered around. Line~3: Modulus (left) and phase (right) of their
    cross-spectrum. Note: A sub-PSF $50\%$ brighter in line 1 has a $225\%$
    brighter autospot in line 2. } \label{fig-simu-j}
\end{figure}

A first ingredient is to compute the correlation between the two diversity
images to access this differential tip/tilt. Since each diversity image is
composed by $\NA$ sub-PSFs (Fig.~\ref{fig-simu-j}~line~1), the image
correlation (Fig.~\ref{fig-simu-j}~line~2) contains $\NA^2$ terms: $\NA$
terms that will be called \textit{autospots} in the following, obtained by the
correlation of the sub-PSF from sub-aperture $n$ in one diversity image by the
sub-PSF from the same aperture $n$ but in the other diversity image. These
autospots lie in the central circled area in Fig.~\ref{fig-simu-j}~line~2.
Outside this circle are $\NA(\NA-1)$ so-called \emph{interspots}, obtained by
the correlation of the sub-PSF from sub-aperture $n$ in one diversity image by
the sub-PSF from sub-aperture $n'\neq n$ from the other diversity image. Those
two kinds of correlation terms are clearly distinguishable by their positions.
Indeed, the interspots result from the differential tip/tilt between the two
involved sub-apertures, which can be large during the first alignment steps.
The autospots positions, near the origin, are not affected by the tip/tilt of
their related aperture, but only by the deterministic diversity-induced
tip/tilt. Fig.~\ref{fig-simu-j}~line~2 shows that the autospots, whose
amplitude scales with $\rho_n^4$ according to Eq.~(\ref{eq-otf-incoh-complete}),
allow the simple estimation of the intensity of each sub-PSF.

The second ingredient is to make these autospots essentially not overlapping. Firstly, using a
sufficiently large diversity (Appendix~\ref{min_div}), so that the sub-PSFs
are shifted from one diversity plane to the other, it is possible to have all
the autospots' cores separated. The small overlapping via the diffraction
rings creates a slight deterministic coupling, which decreases as the
separation increases, but does not prevent the autospots to form a nearly
orthogonal family. Secondly, if the relative tip/tilts between sub-apertures are
large enough with respect to the diversity, autospots are only slightly affected
by the diffraction rings of the interspots. Indeed, the interspots are outside
the circle overplotted on Fig.~\ref{fig-simu-j}~line~2 as soon as the distance (derived as a tip/tilt value in
Appendix~\ref{min_dist}) between two spots is larger than
the diversity-induced shift from one plane to the other. If these two conditions are
fulfilled, each of these autospots can be isolated by projecting the
correlation on a predefined filter matching each autospot. From now on, the
configuration in which the sub-PSFs have the minimal distance between each
other to satisfy the second condition is called \textit{parking position}. 

The third ingredient is to rather compute the FT of this correlation, also
called cross-spectrum, which contains the same information (under a different
form). This cross-spectrum can be computed as the simple product between the
first diversity image FT ($\Ws_1$ from subsection~\ref{deriveOTFPSF}) and the
second diversity image conjugated FT ($\Ws_2^*$). The autospot and interspot FTs will be
respectively called \textit{autopeaks} and \textit{interpeaks}. Although all
the peaks are superimposed in the center of the Fourier plane
(Fig.~\ref{fig-simu-j}, bottom line), the auto/inter
peaks inherit the (near) orthogonality properties of the autospots and
interspots by the Parseval-Plancherel theorem.

\begin{figure}\centering
  \includegraphics[width=\linewidth]{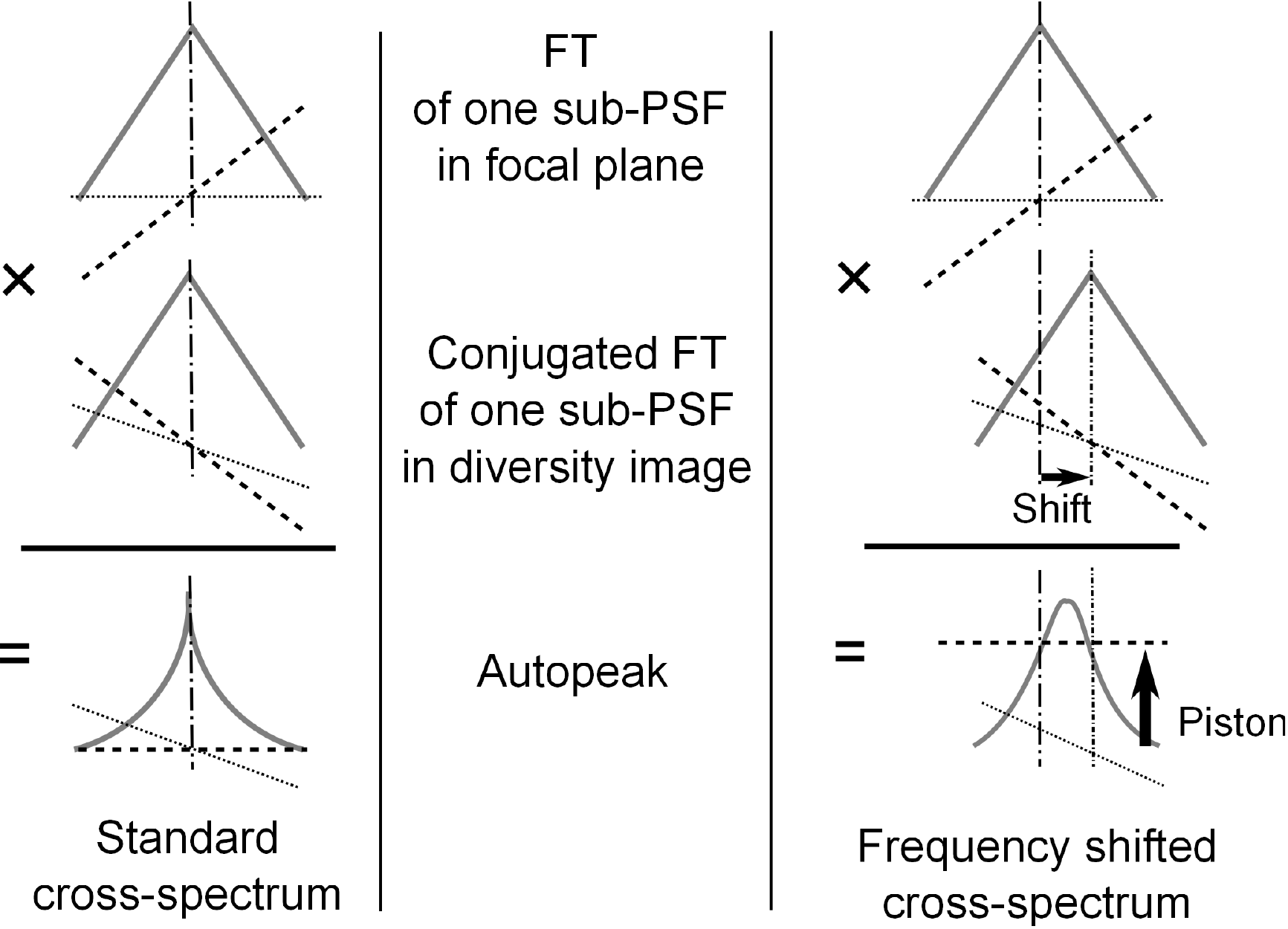}
  \caption{Left column: Standard cross-spectrum (bottom line) as the
    result of the product between one sub-PSF FT from one sub-aperture in the
    first diversity plane (line 1) with the conjugated sub-PSF FT of the same
    sub-aperture in the second diversity plane (line 2). Right column :
    Frequency shifted cross-spectrum (bottom line) as the
    result of the product between one sub-PSF FT from one sub-aperture in the
    first diversity plane (line 1) with the shifted and
    conjugated sub-PSF FT of the same (line 2).
    Each complex term has a a modulus (thick line) and a phase that
    includes two components : the input tip-tilt (dashed line) and the
    diversity (here a pseudo-defocus, in dotted line).}
  \label{fig-principe-j}
\end{figure}

The fourth ingredient of the method is to code the sought tip-tilts in the
autopeaks. In the basic cross-spectrum computation (Fig.~\ref{fig-principe-j},
left) the
contribution of the seeked tip-tilt, in the autopeaks, is cancelled by the
phase conjugation. The idea is to realize that (1) the sought tilt is a phase
ramp in the FT of the sub-PSF, and (2) if we perform a slight shift of the FT
of one sub-PSF, this phase ramp is transformed into itself plus a piston, whose
amplitude is proportional to the input tilt coefficient and the shift offset
(Fig.~\ref{fig-principe-j}, right).
The nice feature of this piston factor is that
it is constant all over the autopeak, and thus can be factored as a
global weighting coefficient, like its amplitude $\rho_n^{4}$. The used operator,
detailed in appendix~\ref{app-fscs}, is hereafter called the
Frequency Shifted Cross-Spectrum (FSCS).

The last ingredient, in the context of a closed-loop control, is to bring the sub-PSFs from a
random scattered state to the parking position. From this parking
position, an open-loop pre-defined offset can then superimpose all the sub-PSFs in the
focal plane. 

This has led to the ELASTIC (Estimation of Large Amplitude Sub-aperture
Tip/tilt from Image Correlation) algorithm, which we now explain with all the
technical details.

\subsection{Derivation of the algorithm}\sublabel{sec-elastic-direct-model}

In practice, the inputs of the algorithm are the diversity images from a
$\NP\times \NP$ pixel detector. The typical width (in pixels) of a sub-PSF is
thus equal to the sampling factor $n_s$ defined from the pixel pitch $p_{pix}$
by:
\begin{equation}
  n_s=\frac{\lambda\,F}{2Rp_{pix}}.
\end{equation}

Note that $n_s=2$ if the sub-PSFs are Shannon-sampled; $n_s$ is greater in
practice for the interferometric PSF to be at least Shannon sampled.
To compute the FSCS without wrapping, images are zero-padded to a width of
$2\NP$ pixels before the computation of the discrete OTFs $\Ws_{d,\Vp}$ by a
Discrete Fourier Transform (DFT). The support width of $\Ws_{d,\Vp}$ is thus
$4\NP/n_s$ pixels. Then, we approximate the discrete OTFs $\Ws_{d,\Vp}$ to a
sampled version of their continuous OTF model $\Ws_d$ derived in
Eq.~(\ref{eq-otf-incoh-complete}). The link between the 2D index $\Vp$ of
$\Ws_{d,\Vp}$ and the continuous reduced coordinates $\Wu$ of
Eq.~(\ref{eq-otf-incoh-complete}) is thus $\Wu=\Vp n_s/\NP$ since the support
half width of $\Ws_d$ is $|\Wu_{max}|=2$. Therefore:
\begin{equation}\label{eq-otf-incoh-complete_discrete}
\Ws_{d,\Vp}=\sum_{n=1}^{\NA}\rho_n^2\Lambda_{n,d}\GP{\frac{n_s\Vp}{\NP}}
   \exp{\GC{j\sum_{k=2}^{3}4\pi a_{kn}\WZ_{k}\GP{\frac{n_s\Vp}{2\NP}}}}.
\end{equation}

Let $\Delta_m$ the operator that performs a shift $\Vd_m$ ($m={2,3}$), with
an amplitude of $\delta$ pixel(s):
\begin{xalignat}{2}\label{eq-def-Delta}
  \Vd_2 & =(\delta,0), & \Vd_3 & = (0,\delta). 
\end{xalignat}

We define the FSCS vector $\Vj_{m,\Vp}$ for each value of $m$ (2 or 3) and
pixel $q$ as:

\begin{align}\label{jim}
    \Vj_{m,\Vp}=&\Ws_{1,\Vp} \times \Delta_m\GC{\Ws_{2,\Vp}^{*}},
\end{align}
where $\cdot^*$ denotes complex conjugation. This computation is illustrated
on the right part of Fig.~\ref{fig-principe-j}. Inserting Eq.~(\ref{eq-otf-incoh-complete_discrete}), and
keeping only the $\NA$ autopeaks as explained in
section~\ref{sec-elastic-principle}, Eq.~(\ref{jim}) becomes:
\begin{align}\label{J-image_auto}
  \Vj_{m,\Vp}=\sum_{n=1}^{\NA} & \rho_n^4\,
   \GP{\Lambda_n\times\Delta_{m}\GC{\Lambda_n^*}}\GP{\frac{n_s\Vp}{\NP}} \\
  & \times \exp\GC{j\sum_{k=2}^34\pi a_{k,n}(\WZ_k-\Delta_m\GC{\WZ_k})\GP{\frac{n_s\Vp}{2\NP}}}.\nonumber
\end{align}

According to Eqs.~(\ref{eq-z1})-(\ref{eq-z3}) from Appendix~\ref{app-zern}: 
\begin{equation} 
(\WZ_k-\Delta_m\GC{\WZ_k})\GP{\frac{n_s\Vp}{2\NP}}=-\frac{n_s\delta}{\NP}\delta_{k,m},
\end{equation}
with $\delta_{k,m}$ the Kronecker delta. 

Thus Eq.~(\ref{J-image_auto}) becomes:
\begin{align}\label{jmm}
 \Vj_{m,\Vp}&=\sum_{n=1}^{\NA}\rho_n^4\exp\GC{-j\frac{4\pi~n_s\delta}{\NP}a_{m,n}}
 \pic_{n,m}\GP{\frac{n_s\Vp}{\NP}},\\
\text{with~} \pic_{n,m}&=\Lambda_n\times\Delta_{m}\GC{\Lambda_n^*}.
\end{align}

Figs.~\ref{fig-illustration} and \ref{fig-simu-j} suggest that the FSCS
computation can be limited to a predefined reduced set of indexes, the OTF central
part, where the autopeaks values are significant. So let us concatenate the
corresponding values of $j_{m,\Vp}$ into a vector $\Vj_m$. Besides, our unknowns
of interest are the tilts $a_{m,n}$, and possibly the transmittances $\rho_n$.
So we group them into a new vector of complex unknowns :
\begin{align}\label{analy-vect}
\Wx_m&= \GC{x_{m,1},...,x_{m,\NA}}^T, & m\in \{2,3\},\\
\mbox{with } x_{m,n}&\eqdef \rho_n^4\exp\GC{-j\frac{4\pi~n_s\delta }{\NP}\,a_{m,n}}, 
& n\in\{1,...,\NA\}.\label{analy-val}
\end{align}

Lastly, Eq.~(\ref{jmm}) is obviously linear in $\Wx_m$, so it can be rewritten in
the final matrix form of the direct model:
\begin{equation}\label{mat_form}
\Vj_m=\VV{C_{m}}\,\Wx_m , 
\end{equation}
where $\VV{C_{m}}$ is a matrix with $2\times\,\NA$ columns, a number of
lines equal to the number of indices kept in $\Vj_{m}$ and the elements of
$\VV{C_{m}}$ are made with the appropriate sampled values of $\pic_{n,m}$.

The generalized inverse $\VV{C_m^\dag}$ of $\VV{C_m}$ can easily be computed
by Singular Value Decomposition. Thus the resolution of the inverse
problem yields the solution $\EST{\Wx_m}$:
\begin{equation}\label{proj}
  \EST{\Wx_m}=\VV{C_m^\dag}\,\Vj_m.
\end{equation}
The seeked tip/tilts and pupil amplitudes are then simply computed as:
\begin{equation}\label{estim}
  \left\{  
  \begin{array}{r@{}l}\displaystyle
    \EST{a}_{m,n}&=\frac{-\NP}{4\pi~n_s\delta}\,\operatorname{Arg}(\EST{\Wx}_{m,n}),\cr
    \EST{\rho}_{m,n}&=|\EST{\Wx}_{m,n}|^{1/4}.
  \end{array}\right. 
\end{equation}

A self-consistency test can check that both amplitude estimations agree:
$\forall~n, \rho_{2,n}\simeq\rho_{3,n}$. It must be noted that $\VV{C_m^\dag}$ can be pre-computed, so that the only
real-time operations are the computation the two (tip and tilt) frequency shifted
cross-spectra from the two (possibly undersampled) diversity images
(Eq.~(\ref{jmm})), then the two
projections (Eq.~(\ref{proj})) and arguments of Eq.~(\ref{estim}). Moreover, for a given sub-aperture
tip/tilt, the sub-PSF shifts (and thus the OTF slopes) are the same for all
wavelengths. The ELASTIC algorithm can thus operate with broadband illumination,
as long as the sub-OTFs $\Lambda_{n,d}$ considered in the direct model are the
polychromatic OTF obtained by averaging over the spectral band the
monochromatic sub-OTFs weighted by the source amplitude and the detector
efficiency. 

\section{Algorithm optimization and performance}\label{sec-optim-perf}
\begin{figure}[!ht]\centering
  \includegraphics[width=0.5\linewidth]{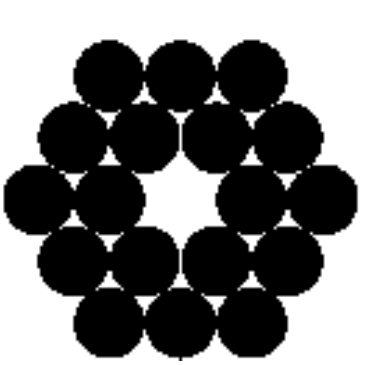} 
  \caption{The $18$ sub-apertures pupil considered for simulations,
    sampled on a 512$\times$512 pixel grid.} \label{JWSTpup}
  \end{figure}

The optimization and performance evaluation performed in this section by
numerical simulations are based on a compact 18~sub-aperture pupil chosen to
mimick the JWST pupil (Fig.~\ref{JWSTpup}), with circular sub-apertures
for compatibility with our software.

\subsection{Definition of the simulation parameters}
Quasi-monochomatic images of size $\NP^2=1024 \times 1024$ pixels are simulated from an
unresolved object. The focal-plane detector is assumed to be
sampled at the Nyquist frequency in the fine phasing mode. Since each sub-aperture diameter is
$1/5$ of the full aperture, the sub-PSFs are oversampled and the sub-aperture
sampling ratio, $n_s$, is $10$. An amplitude of $A_4=0.9$ wave
(Eq.~(\ref{defocus})) is chosen for starters but will be detailed later.

Images are simulated with a total of $3\times 10^5$ photo-electrons,
Poissonian photon noise and a $5$ electrons per pixel read-out-noise.
Algorithm performance is quantified with the estimation of the Root Mean
Square Error (RMSE) of $\NO$ outcomes (with $\NO=50$) defined as:
\begin{equation}\label{eq-def-rmse}
RMSE=\sqrt{\left\langle\frac{1}{\NA}\sum_{m,n}(\widehat{a}_{m,n}-a_{m,n})^2\right\rangle},
\end{equation}
where $\widehat{a}_{m,n}$ and $a_{m,n}$ are respectively the estimated and
introduced aberration coefficients in waves. $\langle\cdot\rangle$ is the average
over the $\NO$ outcomes.

\subsection{Tilt dynamic range and choice of $\delta$}
\sublabel{sec-optim-shift}
The maximum tilt that can be measured is ultimately limited by the detector
field. This maximum tilt, noted $a_{field}$, is given by
Eq.~(\ref{eq-conv-tilt}) assuming a $\pm \NP/2$ shift from the central origin.
\begin{equation}\label{eq-max-field}
  a_{field}=\frac{\NP}{8n_s}.
\end{equation}
Here, $a_{field}=1024/(8\times\,10)\simeq13$ waves. According to Eq.~(\ref{estim}), the
phase of the $\Wx_{m,n}$ coefficients is proportional to the seeked aberration
coefficients $a_{m,n}$ and to the
shift amplitude $\delta$ chosen for the computation of the FSCS. Because the dynamic range of the
$\operatorname{Arg}$ function is only $(-\pi;\pi]$, increasing the shift amplitude
reduces the range since from Eqs.~(\ref{estim}) and~(\ref{eq-max-field}), the
maximal unwrapped estimation is:
\begin{equation}\label{amax}
  a_{max}(\delta)=\frac{2a_{field}}{\delta}=\frac{\NP}{4n_s\delta}.
\end{equation}
Eq.~(\ref{amax}) highlights that the algorithm does not limit the field when
$\delta\leq2$.

\begin{table}[!ht]\centering
  \caption{Variation of field coverage with $\delta$, and corresponding
    maximum aberration coefficient $a_{max}$ in our simulated case.}
  \label{tab-max-tilt}
\begin{tabular}{|c|c|c|c|c|}
\hline
Shift amplitude ($\delta$) & $1$ & $2$ & $3$ & $4$ \\
\hline
Image field coverage ($\%$) & $100$ & $100$ & $66,6$ & $50$ \\
\hline
$a_{max} (wave)$ & $25,6$ & $12,8$ & $8,5$ & $6,4$ \\
\hline
\end{tabular}
\end{table}

\begin{figure}[!ht]
\includegraphics[width=\linewidth]{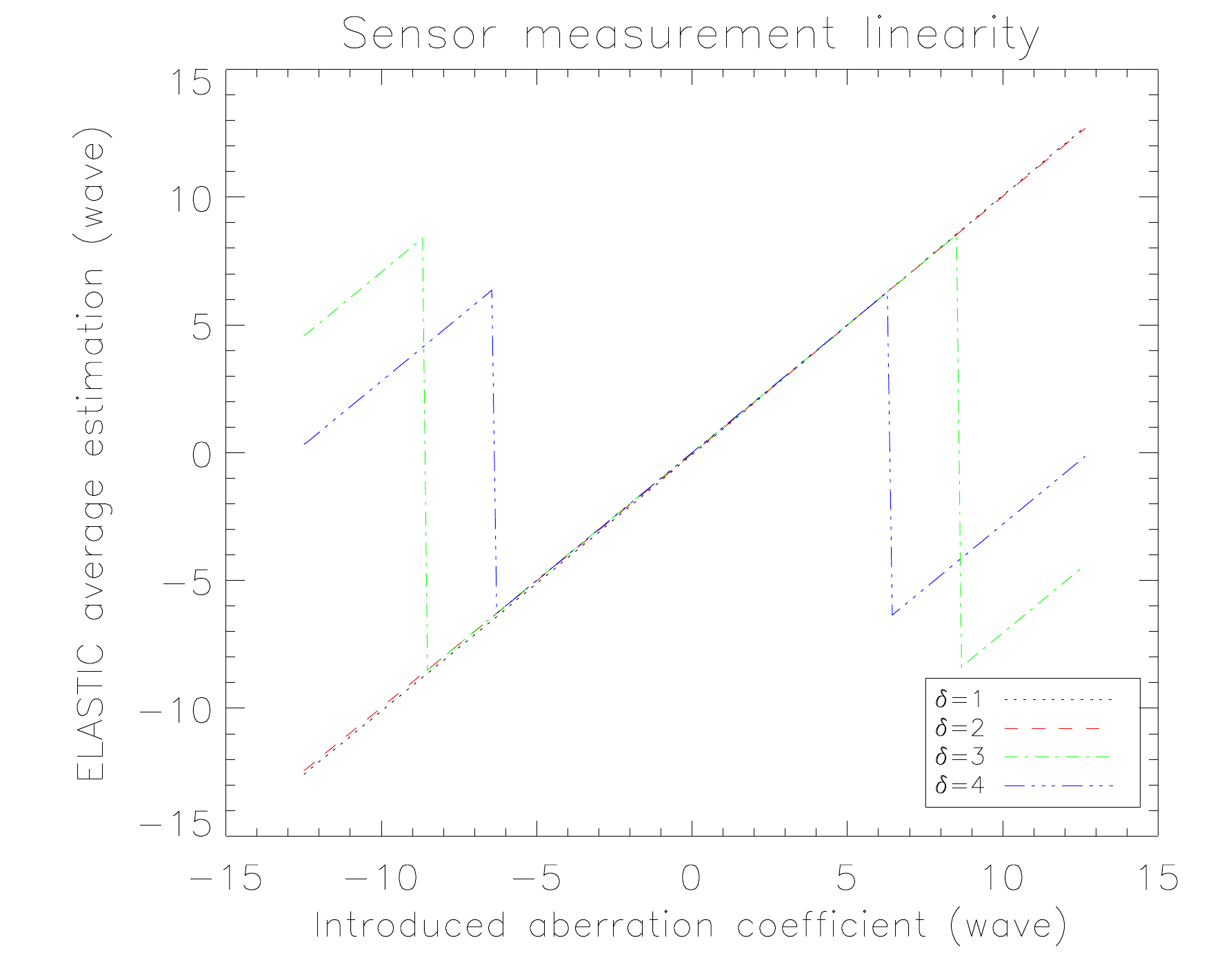}
 \caption{Sensor linearity for various shift amplitudes.}
 \label{shiftlin}
\end{figure}

These limits, given in Table~\ref{tab-max-tilt}, are confirmed by simulation.
Fig.~\ref{shiftlin} presents the ELASTIC estimation versus the introduced
tip aberration for different values of the shift amplitude~$\delta$. A tip
slope is applied over one sub-aperture, the other sub-apertures being randomly
scattered in the field. Each point on the graph is the average of $\NO$ noise
outcomes. A full-field linear response is obtained for a 1 and 2 pixel shifts.
The curves corresponding to 3 and 4 pixel shifts confirm the previously
mentioned phase wrapping, with the theoretical boundaries around $66\%$ or
$50\%$ field coverage, respectively for a 3 or 4~pixel shift.

A 1 or 2~pixel shift are necessary for a maximum field coverage, hence can be
chosen for initial error estimation. Then, as the field covered by the
sub-PSFs decreases, the pixel shift amplitude can be increased to
improve the estimation accuracy as seen in next section.

\subsection{Optimization of the algorithm's parameters} \sublabel{sec-perf-algo}

\begin{figure}[!ht]\centering
  \includegraphics[width=\linewidth]{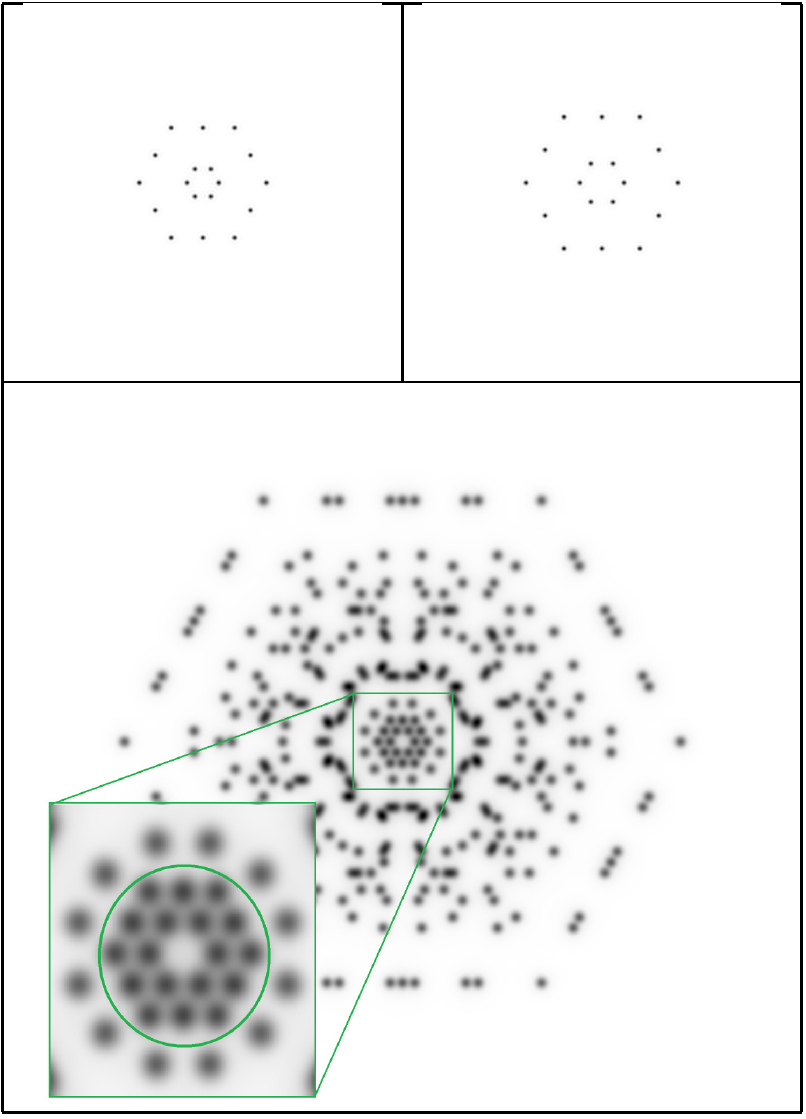}
  \caption{Line 1: Image simulation with
    sub-PSFs in parking positions in the focal image (Left) and in the pseudo
    defocused image (Right). Line 2: Image correlation. Autospots in
    the circle are
    separated and isolated
    from the interspots. } \label{fig-simu-j-18pup}
\end{figure}

The ELASTIC algorithm has two free parameters: the introduced diversity
and the frequency shift amplitude $\delta$. As mentioned in the previous
subsection, it is required to start the alignment with $\delta=2$ if the
sub-PSFs are spread over the whole detector field. The goal here
is to optimize and quantify performance in the ultimate step, when the sub-PSFs are
close to the parking position. Considering
Eqs.~(\ref{analy-val}-\ref{proj}), each $\EST{x}_{m,n}$ value can be written as
$x_{m,n}(1+\Weps_n/x_{m,n})$, with $\Weps_n$ the small noise. Eq.~(\ref{estim}) becomes:
\begin{equation}
\EST{a}_{m,n}\simeq\,a_{m,n}+\frac{\NP}{4\pi n_s\delta}\operatorname{Im}\GP{\frac{\Weps_n}{x_{m,n}}},
\end{equation}
with $\operatorname{Im}(\cdot)$ the imaginary part of a complex number. Because $\Weps_n$ is roughly
independant of $\delta$, it comes that increasing $\delta$
reduces the error on the estimation (if $\NP$ and $n_s$ are fixed). 
Therefore the parking position (depending on the pseudo-defocus amplitude) needs to be
tightly packed for two reasons. Firstly, it
allows to increase $\delta$ hence to improve the estimation accuracy.
Secondly, it will minimize errors due to the uncertainty of segment
displacements (for the open loop superimposition after ELASTIC).
Fig.~\ref{fig-simu-j-18pup} presents the parking position chosen for the
sub-PSFs. The positions of the sub-PSFs from the central corona were
determined thanks to Appendix~\ref{min_dist}. Then, the sub-PSFs from the
external corona were positioned so that the interspots do not overlap the
autospots after image correlation (Fig.~\ref{fig-simu-j-18pup}, line~2).

\begin{figure}
\includegraphics[width=\linewidth]{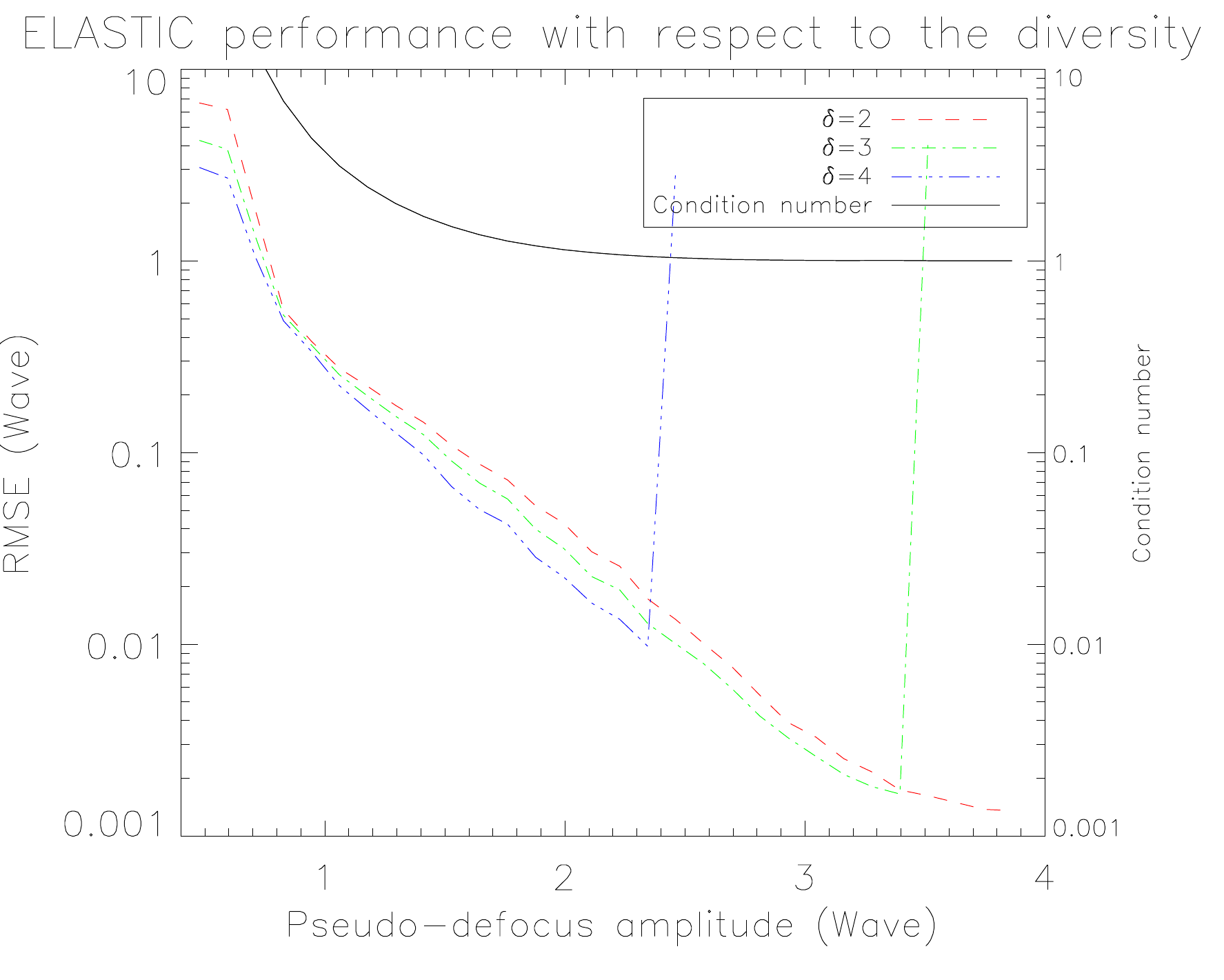}
\caption{Estimation RMSE vs diversity for various shifts $\delta$.}
\label{RMSE-div}
\end{figure}

Fig.~\ref{RMSE-div} presents the evolution of the ELASTIC estimation RMSE
versus the pseudo-defocus amplitude $A_4$, for different values of the shift amplitude
$\delta$. It is clearly evidenced that ELASTIC estimation RMSE decreases when the
pseudo-defocus amplitude and the pixel shift amplitude are increased. The
algorithm performance is better than $\lambda/8$ for a pseudo-defocus larger
than $0.8$ to $1$ wave. This is in agreement with the typical value of $0.9$
wave given by Eq.~(\ref{defocus}). For a larger diversity amplitude the RMSE
estimation goes down to less than $\lambda/500$ for a $2$ or $3$~pixel shift.
Indeed, when the diversity amplitude increases, the interspots are shifted further from the autospots.
Hence their influence on the autospots decreases then the estimation's
accuracy is better.
The limit to these performances is that the
diversity amplitude must be chosen so that the sub-PSFs do not slip out of the
unwrapped field (settled by $\delta$). This limitation is not fulfilled when
the pseudo defocus amplitude is
larger than 2.3~waves and 3.3~waves for respectively a 4 and 3~pixel shift
amplitude: the RMSE increases. There is then a trade off between the choice of
$\delta$ and the pseudo-defocus amplitude. 
The evolution of the condition number of $\VV{C_m}$ is also
plotted. We see that it is indeed a good indicator of the matrix inversion's sensivity to noise. As the ratio
between the highest and the smallest singular values, it can help one qualify
the ability to separate autospots as a function of the pseudo-defocus
amplitude. The condition number decreases when the
pseudo-defocus amplitude increases, and reaches an asymptote at 1. Indeed a
large diversity improves the autospots separation. Therefore, the singular modes are well distinguishable
for the matrix inversion. The condition number is the same for each shift
amplitude case. Indeed the autospot separation in the correlation only depends on the diversity,
not on the introduced frequency shift. Hence, increasing the latter makes the estimation
more accurate. 

\subsection{Robustness to noise} \sublabel{sec-noise-propag}

\begin{figure}[!ht]
\includegraphics[width=\linewidth]{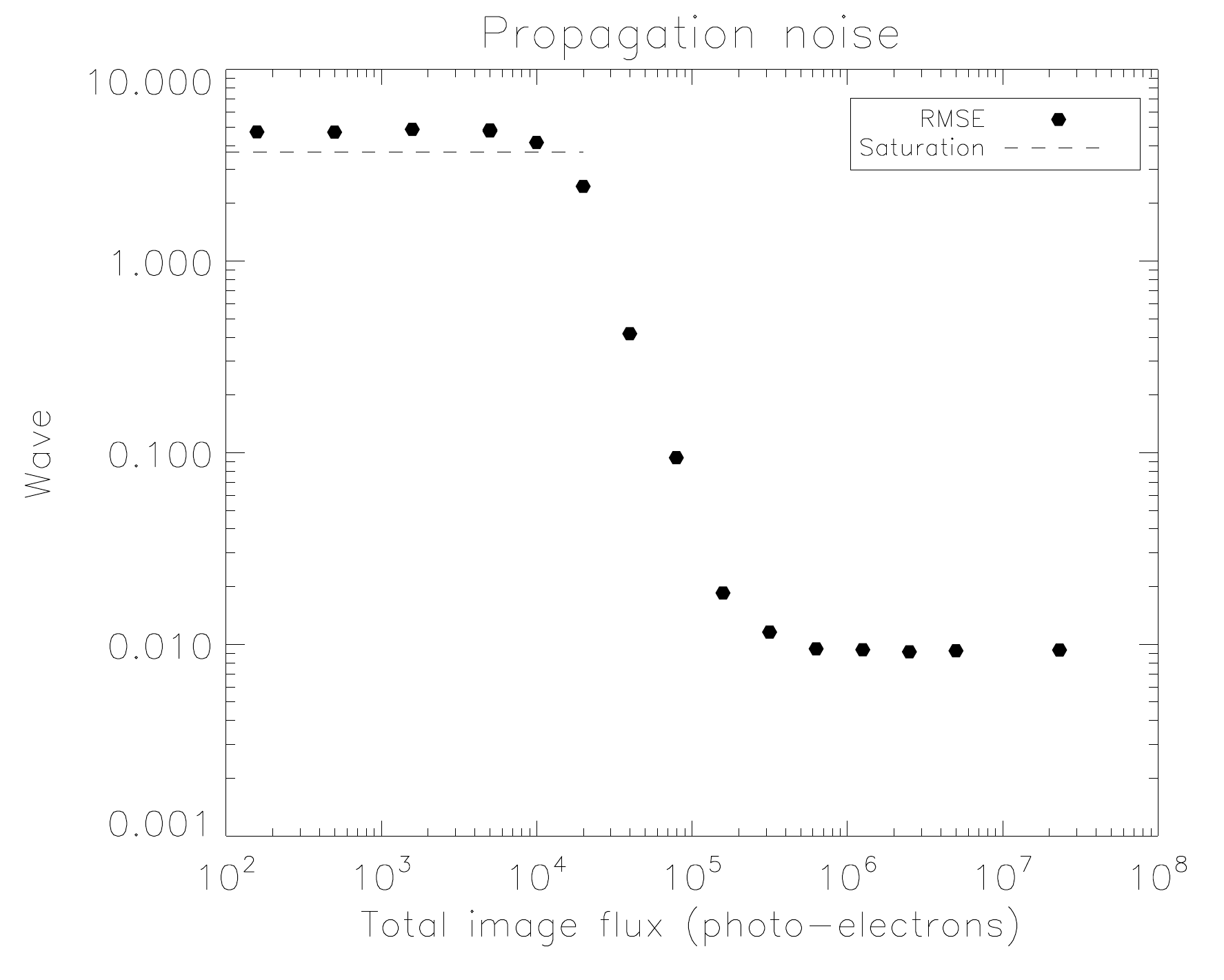}
 \caption{Wavefront error in waves as a function of the image flux for
   $\delta=4$, and for a 18 sub-aperture instrument.}
 \label{propagnoise}
\end{figure}

In order to have the best accuracy with the largest frequency shift amplitude
and a relatively small defocus amplitude,
considering the sub-PSFs in parking position, a $2.2$~waves pseudo-defocus and a $4$~pixel shift
amplitude are chosen for the evaluation of noise propagation. $\NO$ noise
outcomes are simulated for different image brightnesses, quantified by
$\NPH$, the total number of photo-electrons.  Fig.~\ref{propagnoise}
presents the estimation's RMSE as a function of the
flux.

Several regimes are to be noticed on Fig.~\ref{propagnoise}. Below
$10^4$\,photo-electrons (low illumination), the estimated RMSE is
constant. In this case, the distribution of
$\operatorname{Arg}(\EST{\Wx}_{m,n})$ behaves as a uniform noise between
$-\pi$ and $\pi$. Such a distribution has a theoretical standard deviation of
$\pi/\sqrt{3}$. Using Eq.~(\ref{estim}), we deduce that the theoretical
saturation of the RMSE is $\NP/(4~n_s\delta\sqrt{3})=a_{max}/\sqrt{3}$. The numerical value
of the saturation for our setup is around $4$~waves, and matches with the
saturation observed in the simulation.
 
A second regime is for a total image flux greater than
$10^5$\,photo-electrons: the estimation RMSE
is saturated.  This last limitation is the algorithm's bias. The bias can be
explained by the small influence of the interpeaks on the autopeaks.
However, the RMSE is down to less than $\lambda/100$, which is already much
better than needed to enter the fine phasing mode. 

These results confirm the capacity of ELASTIC to bring a
misaligned multi-aperture telescope to a configuration state where the wavefront errors are
$\lambda/100$ allowing the fine phasing algorithms to operate.

\subsection{Robustness to phase diversity error} 
An interesting feature of the ELASTIC algorithm is its small sensitivity to the diversity error.
This can be understood in the focal domain: if the diversity used in the numerical model does not match exactly the actual optical diversity, 
then the autospots in the frequency-shifted cross-correlation (cf appendix~\ref{app-fscs}) do not perfectly overlap the associated data in the FT of the $\VV{C_m^\dag}$ vectors. 
The estimated intensity will be affected by the partial overlap. But since the piston information induced by the seeked tip/tilt is constant and not  degraded over the autospot, its extraction is not much affected (neglecting the overlapping of the spots associated to different sub-apertures).
This is confirmed by simulations: the estimation error is less than $\lambda/8$ as long as the diversity-induced 
tip or tilt error is less than $\lambda/2.5$ on each sub-aperture \cite{vievard2017developpement}.

\subsection{Adaptation to atmospheric turbulence}
In the case of a ground-based telescope, the atmospheric turbulence induces (among others) tip-tilt disturbances on the sub-PSFs. 
A solution to average the  random atmospheric tip/tilt is to record long-exposure images, leading to sub-PSFs of size $\lambda/r_0$  (instead of $\lambda/D$) with $r_0$ the Fried parameter \cite{Roddier-88}.
Thus, the ELASTIC algorithm can be adapted by changing the  photometric peaks $\Lambda_{n,d}$ by narrowed versions because frequencies above $r_0/\lambda$ (instead of $D/\lambda$) are lost.
All the remaining processing can be performed, leading to a fraction of
$\lambda/r_0$ precision (instead of a  fraction  of $\lambda/D$). This
accuracy should ensure that we enter the capture range of usual wavefront sensors over all the sub-apertures simultaneously, in order to operate an adaptive optics system.

\section{Experimental validation}\label{exp-valid}
In the following, sub-PSFs of a multi-aperture telescope are brought to the
parking position thanks to the ELASTIC algorithm, and then superimposed.

\subsection{Implementation of ELASTIC}
In order to test focal plane wavefront sensors, Onera built a dedicated bench
called BRISE \cite{Cassaing-p-06b}.

\begin{figure}[!hb]\centering
\includegraphics[width=\linewidth]{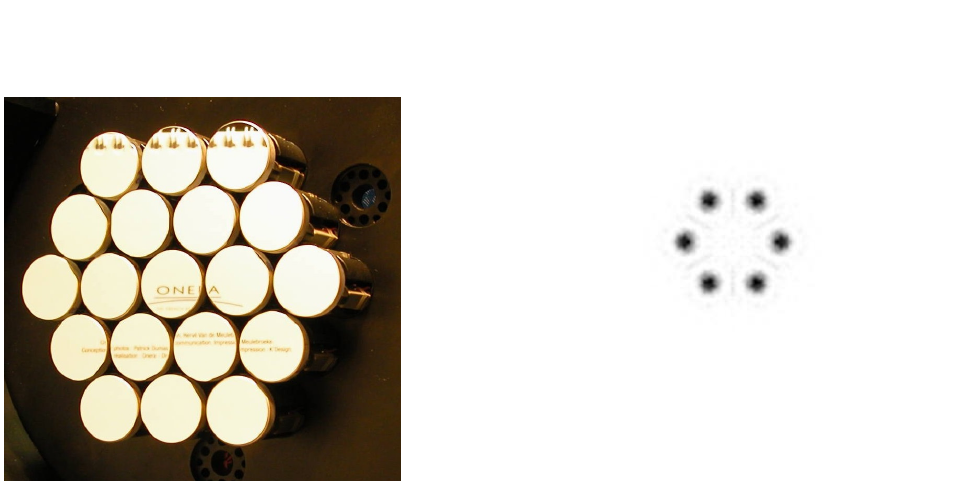}
\caption{Left: the active segmented mirror with 19 sub-apertures. Right:
  Simulation of the parking position in the
  focal plane with the 6 sub-apertures used (first corona).}
\label{mirror}
\end{figure}

A fibered single-mode laser diode operating near $635$\,nm is used as an
unresolved source. It is collimated and provides an object at infinity to a plane
segmented mirror with nineteen sub-apertures (Fig. \ref{mirror}~left). To
introduce and correct piston/tip/tilt perturbations, each of these mirrors is
supported by three piezoelectric actuators. These have no internal feedback
control and suffer from hysteresis.  In addition, the external corona is
currently not fully functional, thus the validation is performed over the $6$ sub-apertures
of the first corona. Downstream, a phase diversity module is used to
simultaneously form a focused and a defocused image of the object on a
$1300~\times~1000$\,pixels camera, from which we extract two $512~\times~512$
images. 

Because the active mounts are not perfectly deterministic, it is not possible
to perform a single-step good correction from a randomly misaligned state.
Therefore, a closed-loop sequence is performed to align the mirror. According
to Eq.~(\ref{defocus}), the minimal required diversity for ELASTIC
is $0.94\lambda$. Measurements are then performed with a $1\lambda$
pseudo-defocus temporally applied on the
512$\times$512\,pixel focal image. To have a full-field tip/tilt estimation,
the FSCS is computed with a 2\,pixel shift.  

\subsection{Loop closure on an unresolved source}
\begin{figure}[!ht]\centering
  \includegraphics[width=\linewidth]{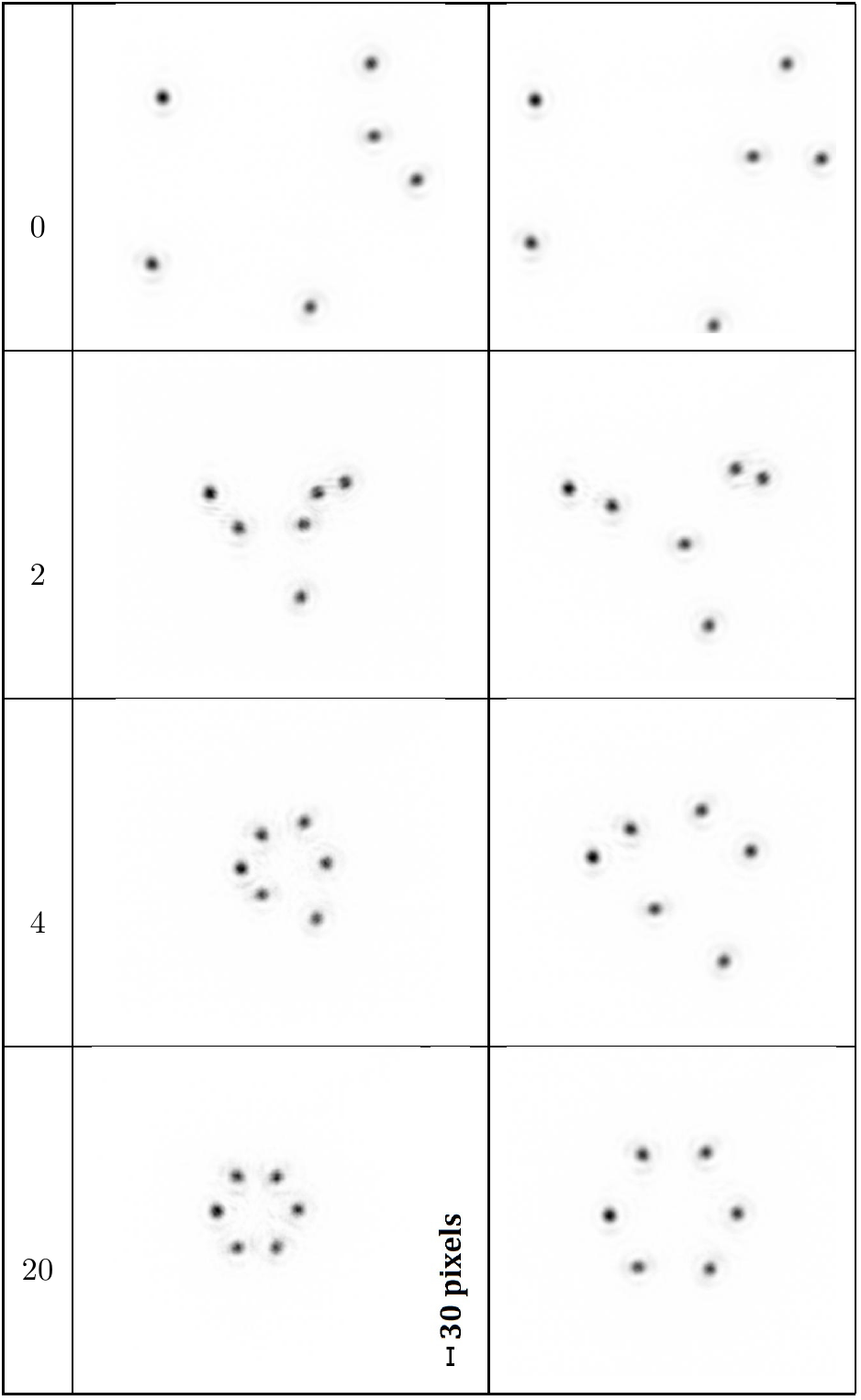}
  \caption{Closed-loop sequence up to the parking position with a $\delta$
  amplitude of 2, and a total flux of 3e7 photo-electrons per image.  Columns are
  iteration number, focal and pseudo-defocused images.}
\label{close-loop}
\end{figure}

\begin{figure}[!ht]
	\centering
	\includegraphics[width=\linewidth]{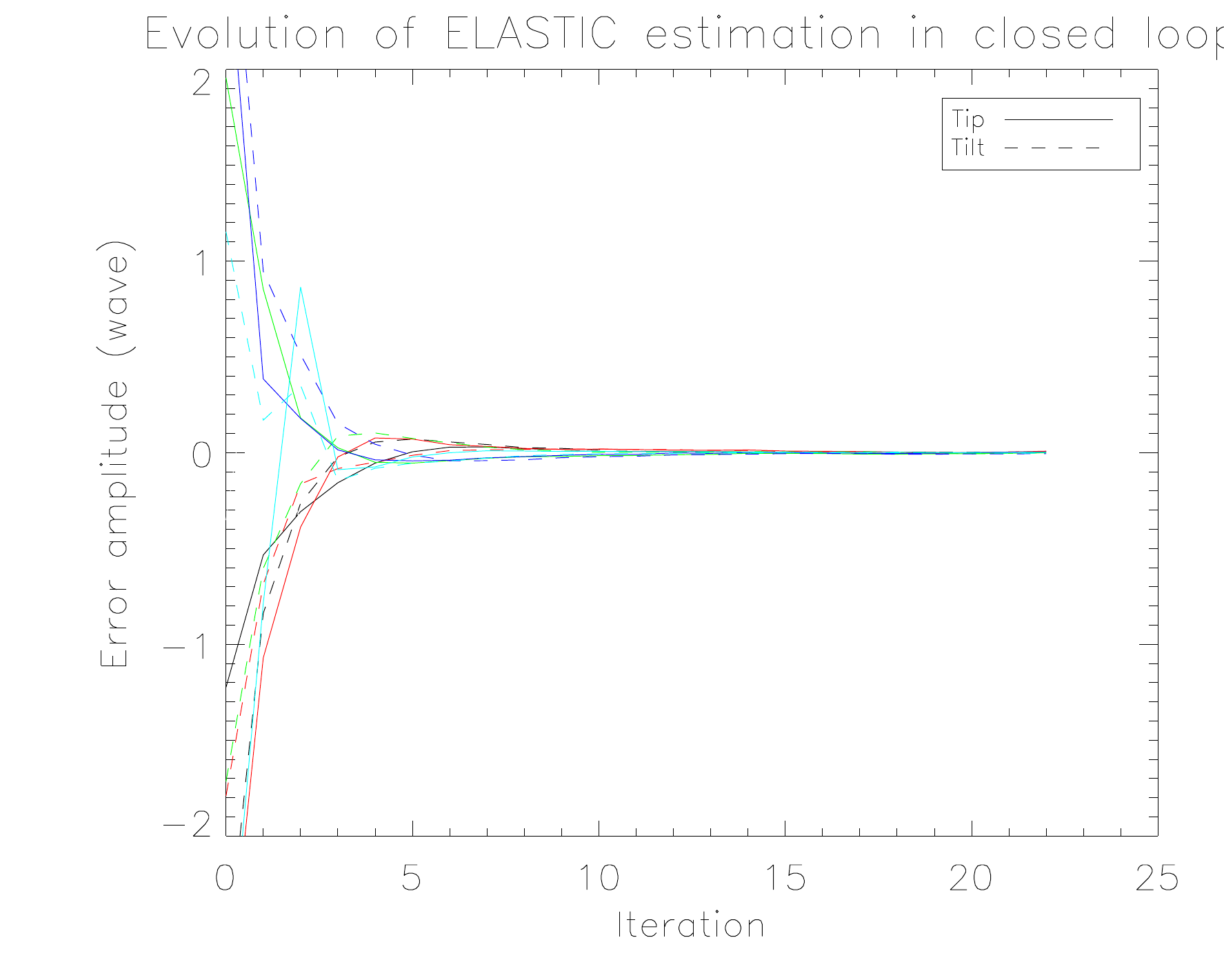}
	\caption{Evolution of Tip/tilt estimation versus the iteration number during
		the loop.}
	\label{tiptiltampl}
\end{figure}

The focal and pseudo-defocused images of an unresolved source are shown on
Fig.~\ref{close-loop}~line~1. They include random tip/tilt errors on each sub-aperture.
The aim is to obtain in the focal plane the same simulated parking position as
Fig.~\ref{mirror}~right. For the closed loop, we use an integrator with a gain
control of $0.5$. Lines~2 and 3 of Fig.~\ref{close-loop} show iterations 2 and 4 of the loop.
Line~4 is the position obtained at iteration~20, close to the expected
parking.  

Fig.~\ref{tiptiltampl} presents the experimental evolution of the estimated tip/tilt errors
during the closed-loop, for each iteration. The error is defined as the relative
distance to the parking position. At first iteration, some tip/tilt errors are
larger than $2~\lambda$. As from iteration~5, the estimated error is mainly
less than $\lambda/2$.  Moreover, the closed-loop control is stable as from
around iteration~7 and keeps the sub-PSFs in the parking position until the end with an
estimated error less than $\lambda/10$. It is to be noted that the estimation
can be biased if two sub-PSFs come too close (iteration 2), but the error
always converge to zero thanks to the closed-loop control.

It can be seen on iteration~20 that the sub-PSFs are not
perfectly placed (they do not form the exact pre-defined parking). This is
mainly due to the temporal modulation over each sub-aperture. Indeed, in open
loop, if the diversity offset is consecutively switched on and off, the
sub-PSFs do not go back to the exact same
position they had before the offset. Hence, the instrumental limitations (plus
a possible error in the telescope modeling and environment potential
perturbations) can explain this visible positionning error on iteration~20.
Nevertheless we can conclude that ELASTIC was succesfully implemented on BRISE bench, and was able
to bring the telescope from a randomly misaligned configuration to a stable
parking position. The error of the estimation, due
to the above mentioned limitations, is now to be quantified.

\subsection{Noise evaluation}
\begin{figure}[!ht]
\centering
\includegraphics[width=\linewidth]{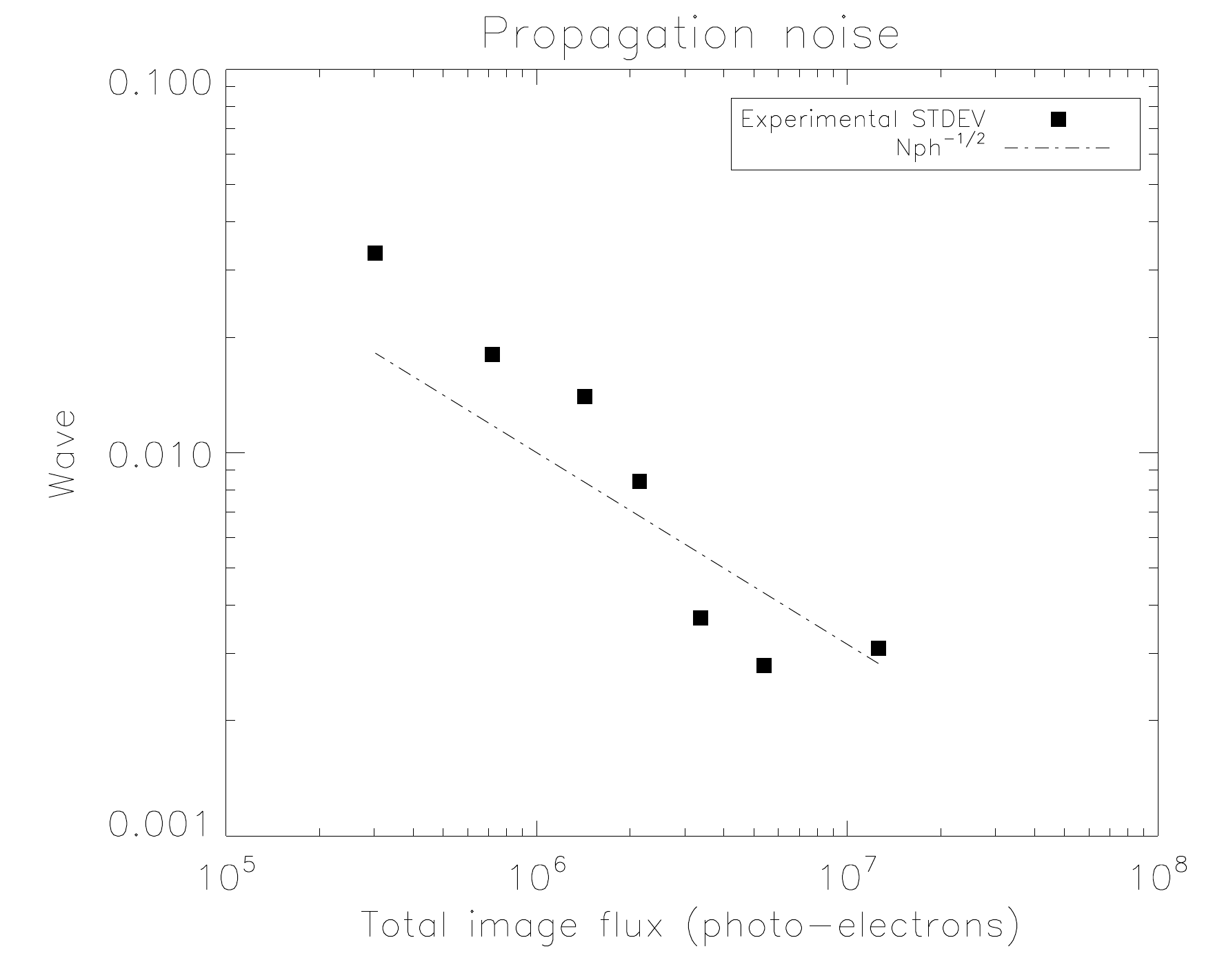}
\caption{Experimental estimation root mean square error for different
  illumination conditions. Each point is computed with 50 outcomes.}
\label{expprop}
\end{figure}

To experimentaly evaluate the estimation standard deviation for several illumination
conditions, the sub-PSFs are set in a fixed configuration (e.g. the
previously mentioned parking position) and $\NO$~open-loop diversity
pairs of images are taken to estimate tip/tilt. Fig.~\ref{expprop} plots the
estimated standard deviation for various illumination values (defined as the
total flux in each image).

Results show that the estimation's standard deviation  is less than
$\lambda/100$ as from a $10^6$ photo-electrons illumination, and decreases following a  $1/\sqrt{\NPH}$ slope. This
highlights the stability of the estimation for reasonable flux conditions.

Despite the instrumental limitations (turbulence from the camera fan,
hysteresis, possible calibration errors) that remain, the stability of the
estimation is better than the capture range of the fine algorithms.

\subsection{Final superimposition}
\begin{figure}[!ht]
  \centering
  \includegraphics[width=\linewidth]{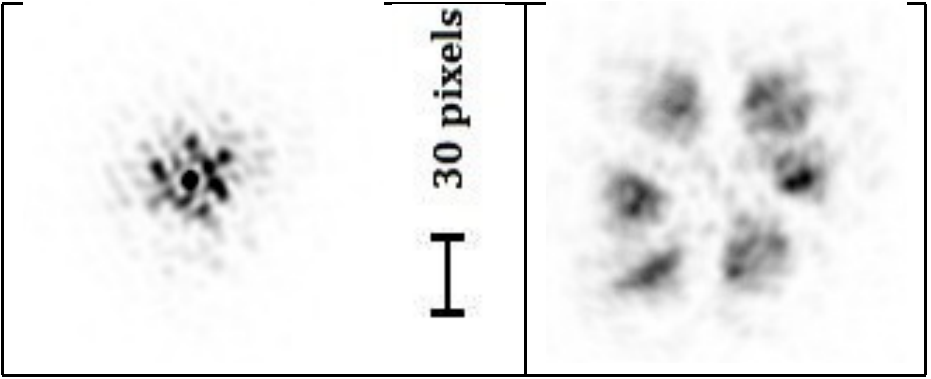}
  \caption{Zoom of focal (left) and pseudo-defocused (right) PSF after final
    superimposition ($128\times\,128$ pixel images).}  \label{End}
\end{figure}

After the closed-loop organisation of the sub-PSFs in parking position, an
open-loop offset command is applied to finish the alignement (Fig.~\ref{End}).
In the focal plane, interference fringes are distiguishable. Eventhough the
large piston errors were not corrected, they are smaller than the coherence
length of the single-mode laser used as an unresolved source. 
Without hysteresis or calibrations errors, simulations showed in Section~\ref{sec-noise-propag} that a $\lambda/100$
precision could be reached for high illumination, in the case of a 18~sub-aperture instrument. Experimentally, the result presented
on Figure~\ref{End} shows that in the case of a 6~sub-aperture instrument, the open-loop offset command 
confines all sub-PSFs in an area which is less than $30$~pixels FWHM. 
Thus the sub-PSFs dispersion after superimposition is +/- 7~pixels peak-valley
around the center, leading to an estimation of residual tip/tilt error smaller than $\lambda/8$ RMS, according to Appendix~\ref{TT-coeff} since
the FWHM of a single sub-PSF is $16$~pixels ($n_s=16$ on our bench).

\begin{figure}[!ht]
  \centering
  \includegraphics[width=\linewidth]{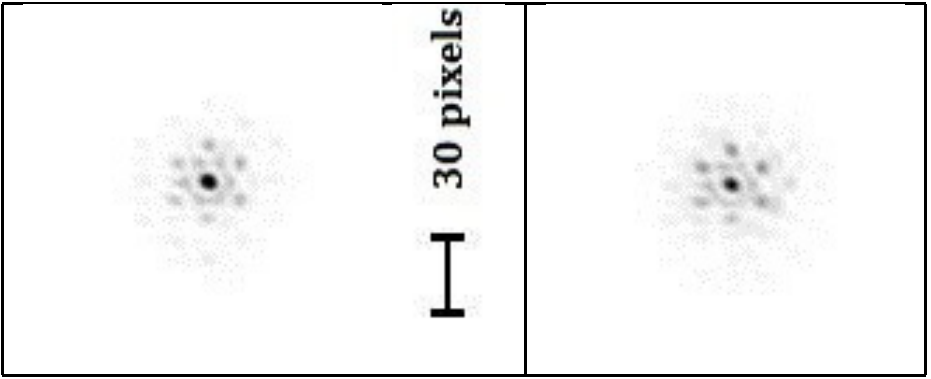}
  \caption{Focal (left) and defocused (right) PSFs after phase diversity fine
    phasing algorithm ($128\times\,128$ pixel images).}  \label{End_}
\end{figure}

This is confirmed by Fig.~\ref{End_}, which shows the result of a cophasing by phase diversity~\cite{Gonsalves-PhaseXretrieval_diversity,Mugnier20061} after
superimposition by ELASTIC. Not only the sub-PSFs are all in the capture range
of the fine phasing algorithm, but the width of the cophased pattern (2~pixel central core but around 30~pixel enveloppe) is
comparable to the width of the aligned state after ELASTIC.

\section{Conclusion}

To estimate the tip/tilts (and pupil transmission amplitude) over each sub-aperture of a
multi-aperture telescope, we introduced the ELASTIC method whose main
features are to provide a polychromatic large amplitude estimation (up to the full camera
field) with a simple hardware (only two \emph{images} of a point
source near the focal plane) and a closed-form unsupervised computation with small computing
cost.  Its typical use is to bring the sub-apertures from any distorted state
up to a sufficiently aligned state (with residues much smaller than half the
sub-PSF width, to enable their interference) to scan for pistons and
ultimately enter the fine-phasing mode. 
Such a source (unresolved by each sub-aperture) can easily be found in a
stellar field. For applications such as Earth observation from space, the
ELASTIC algorithm may be extended to be insensitive to the object phase;
however, it can be expected that large amplitude alignment is not frequent and
can be operated by pointing the telescope at a star to operate  ELASTIC with
an unresolved source.

The main requirement of ELASTIC is to acquire two images, including
known local tip/tilts offsets over the sub-apertures, so that each sub-PSF is
uniquely identified by its position shift.  These offsets need to be large
enough to allow a clear separation between the sub-PSFs, hence are larger than
those required by \emph{phase diversity} in the fine phasing mode, which needs
the sub-PSFs to remain superimposed in all diversity planes to interfere.
This \emph{geometric diversity} can be implemented by inserting a global aberration
between the images (e.\,g. focal/defocused images) or by
introducing a temporal modulation with the sub-aperture
correction actuators themselves. This is also simpler since telecentricity is
not required, and more efficient as only the
tip-tilt part of the defocus is used by the algorithm, whereas higher-order
modes only degrade the sub-PSFs. 


ELASTIC is based on the computation of a frequency-shifted cross-spectrum, an
operator detailed in appendix~\ref{app-fscs} introduced to simply invert the
diversity process. We showed by means of numerical simulations with a JWST-like pupil how to optimize the free
parameters (the diversity and the frequency shift) and estimated the limiting
magnitude around $5\times10^{3}$~photo-electons/sub-aperture/frame with
$1024\times~1024$ pixels images. Lastly, we performed an experimental validation
that demonstrated the closed-loop alignment of a 6~aperture segmented mirror using
a tip/tilt temporal diversity on the segments and
$16\times~10^{4}$~photo-electrons/sub-ap/frame with
$512\times~512$ pixels images.

Experience shows that the ELASTIC procedure is robust and is
now regularly used to drive BRISE into the fine phasing mode. A short-term
perspective is then to
interface ELASTIC with a real-time fine phasing algorithm such as the one
proposed in~\cite{Mocoeur:09,vievard2016real}.

\paragraph{Fundings}
This research was partly funded by ONERA's internal research project VASCO ;
Thales Alenia Space co-funded S.~Vievard's PhD thesis; the BRISE bench was funded by
French DGA 

\paragraph{Acknowledgements}
The authors would like to thank B.~Denolle, A.~Grabowski and
C.~Perrot for their early work on the algorithm and the experiments;  J.~Montri
for the BRISE computer interface; and J.-P.~Amans from GEPI (Galaxies, Etoiles,
Physique et Instrumentation) of Observatoire de Paris-Meudon, for the design and
manufacturing of the segmented mirror.

\appendix
\renewcommand{\thesubsection}{\arabic{subsection}}
\makeatletter
\renewcommand{\sublabel}[1]{\def\@currentlabel{\Alph{section}\arabic{subsection}}\label{#1}}
\makeatother

\section{Appendix}
\renewcommand{\theequation}{A\arabic{equation}}
\subsection{Definition of the Zernike modes}\sublabel{app-zern}
This appendix recalls the Zernike polynomials \cite{Noll-ZernikeXpoly}: a basis of
orthonormal modes $Z_{k}$ with normalized coordinates
$\Wu_n=(u_n,v_n)$, where $|\Wu_n|\leq~1$. The first 4 modes we will use are:
\begin{xalignat}2
   Z_1(\Wu_n)&= 1,      & &\text{(piston)}\label{eq-z1}\\
   Z_2(\Wu_n) &= 2u_n, & &\text{(tip)}\label{eq-z2}\\
   Z_3(\Wu_n)&= 2v_n, & &\text{(tilt)}\label{eq-z3}\\
   Z_4(\Wu_n)&= \sqrt3(2\Wu_n^2-1). &&\text{(defocus)}\label{eq-z4}
\end{xalignat}

\subsection{From global to local aberrations}
\sublabel{app-glo2loc}

The global position $\Wr$ of any point in the pupil can be linked to a
local coordinates $\Wr_n$ attached to each sub-aperture:
\begin{equation}\label{eq-glob2loc-pup}
  \Wr= \Wr_n+R\Wc_n.
\end{equation}
with $\Wc_n=(c_{x,n},c_{y,n})$ the $n$\textsuperscript{th} sub-aperture
center normalized by $R$. We define $B$ as the global aperture diameter, and:
\begin{xalignat}2 
 \Wu_n\eqdef\frac{\Wr_n}{R}, && \Wu '\eqdef \frac{\Wr}{B/2},
\end{xalignat}
with $\Wu_n$ the reduced coordinates over the sub-aperture and $\Wu
'$ the reduced coordinates over the global aperture. Therefore:
\begin{equation}
\Wu '= \frac{2R}{B}\,(\Wu_n + \Wc_n).
\end{equation}

Then, inserting in Eq.~(\ref{eq-z4}), a defocus can be decomposed as:
\begin{align}
  Z_4(\Wu ')=&\sqrt3 \GC{2\frac{4R^2}{B^2}\,\GP{\Wu^{2}_{n}
      +2\,\Wu_n.\Wc_n+\Wc_n^2}-1} 
  \\ =&\frac{4R^2}{B^2}\,Z_4(\Wu_n) +2\sqrt3\frac{2R^2}{B}
       \GC{\frac{c_{x,n}}{B/2}Z_2(\Wu_n)+\frac{c_{y,n}}{B/2}Z_3(\Wu_n)} \nonumber \\
  &+\sqrt3\GC{\frac{2R^2}{B^2/4}\,\Wc_n^2+\frac{4R^2}{B^2}-1}. 
\end{align}

Therefore, the tip (or tilt) induced by a defocus on the $n^{th}$-subaperture is:
\begin{equation}\label{eq-defoc2ai}
  a_{2,n}=2\sqrt3\frac{4R^2}{B^2}c_{x,n} A_4.
\end{equation}

\subsection{Computation of the tip/tilt coefficients}\sublabel{TT-coeff}
A tilt disturbance can be expressed either by an angle $\theta$ in the object
space, by a $k_p$ pixel shift over the detector (with a $p_{pix}$ pixel pitch)
or by a $a_m$ coefficient, in waves, linked by:
\begin{xalignat}3  \label{eq-conv-tilt}
  \theta&=\frac{k_pp_{pix}}{F}=\frac{2a_m\lambda}{R} &&\Rightarrow&
   a_m = \frac{R\theta}{2\lambda}=k_p\frac{R p_{pix}}{2 \lambda F}=\frac{k_p}{4n_s},
\end{xalignat}
since from Eq.~(\ref{eq-z2}) the peak-to-valley amplitude of $Z_2$ is~4.
A consequence from Eq.(\ref{eq-conv-tilt}) is that the FWHM of a sub-PSF corresponds to as tilt variation of
$\Delta\,a_2=1/4$ since the sub-PSF's angular FWMH is $\Delta\theta=\lambda/2R$.

\subsection{Typical diversity for ELASTIC}\sublabel{min_div}
The goal of this appendix is to compute the amount of defocus to have the
distance between the autospots just equal to their diameter. $P$ is defined as
the distance between two aperture centers.
It can be computed that the autocorrelation of an Airy pattern (which has an angular FWHM
of $\lambda/(2R)$ rad) has an angular FWHM of $(2\lambda)/(3R)$ rad or 1/3 of wave from Eq.~(\ref{eq-conv-tilt}). In addition, the separation
between two adjacent sub-apertures with a defocus of amplitude $A_4$ (RMS
value over the global full aperture) is from Eq.~(\ref{eq-defoc2ai}) $\Delta
a_m=2\sqrt3\frac{4R^2}{B^2}\,PA_4$.
Equalizing this last expression with 1/3 gives the typical defocus value:
\begin{equation}\label{defocus}
 A_4=\frac{1}{6\sqrt3}\,\frac{B^2}{4R^2}\frac1P.
\end{equation}

\subsection{Minimal distance between the sub-PSFs}\sublabel{min_dist}
The minimal distance between the sub-PSFs for the interspots not to pollute
the signal is now defined, in the case of a single corona aperture. The
separation between autospots from extremal apertures with a defocus of
amplitude $A_4$ (RMS value over the full aperture) is $\Delta
a_m=2\sqrt3\frac{4R^2}{B^2}(B-2R)A_4$ from Eq.~(\ref{eq-defoc2ai}).  Since
the sub-PSF FWHM corresponds to a $1/4$ tilt from App.~(\ref{TT-coeff}), the minimal separation
is:
\begin{equation} 
\Delta a_m=2\sqrt3\frac{4R^2}{B^2}(B-2R)A_4 + \frac14.
\label{parkingformula}
\end{equation}

\subsection{The Frequency Shifted Cross-Spectrum (FSCS)} \sublabel{app-fscs} 

This appendix considers the various ways to correlate two signals $\Vi_1$
and $\Vi_2$, which here would stand for the two diversity PSFs.

Authors agree to define their cross-correlation or inter-correlation as
$\Vi_{1,2}=\Vi_1\,\otimes\,\Vi_2$ (cf. Eq.~\ref{autocorr}) \cite{goodman2005introduction}.

The cross-spectrum is sometimes defined as the FT ($\mathcal{F}$) of the cross-correlation:
$\mathcal{F}\GC{\Vi_{1,2}}=\mathcal{F}\GC{\Vi_1}\times\mathcal{F}\GC{\Vi_2}^*=\Ws_1\times \Ws_2^*$ and, in imaging
through turbulence
\cite{Fontanella-knoxThompson-josaa87,Roddier-88,roggemann-imagingThroughTurb-96},
as $\Ws_{1,1\V{\delta}}=\moy{\Ws_1\,\Delta_m\Ws_1^*}$ where the average
is performed over turbulent phase outcomes and $\Delta_m$ is the shift
operator by a small increment
$\V{\delta}$ defined in Eq.~(\ref{eq-def-Delta}), used by the Knox-Thompson method to estimate the object phase
from turbulence-degraded images \cite{knox-imageRecoveryAtm-apj74}.  This
cross-spectrum only holds for a stack of single images.

The product used by Eq.~(\ref{jim}), defined as
$\Ws_{1,2\V{\delta}}=\Ws_1\,\Delta_m\Ws_2^*$, is a combination of these
two products: it is performed in the spectral domain, between two different
signals, with a small shift introduced to access some phase information as
detailed by Fig.~\ref{fig-principe-j}. To the best of our knowledge, it
has not been introduced before so we name it the Frequency Shifted
Cross-Spectrum (FSCS).

The FSCS can be equivalently understood in the direct domain from
Fig.~\ref{fig-simu-j}: the spectral shift of $\Ws_2$ is equivalent to
the multiplication of $\Vi_2$ by a phase slope. In the real image of
Fig.~\ref{fig-simu-j}, line~1, right, each sub-PSF would inherit a
phase, proportional to its $x$-coordinate. After what can be called a
frequency-shifted cross-correlation (the FT of the FSCS), the \emph{positions}
of the autospots are still given by the \emph{differential positions} of their
related sub-PSFs in the
two images as illustrated by Fig.~\ref{fig-simu-j}, line~2, but their
\emph{value} would have a phase given by the associated phase (thus
\emph{absolute position}) in the sole $\Vi_2$ image.


\begin{thebibliography}{10}
\newcommand{\enquote}[1]{``#1''}

\bibitem{VLTI-url}
\url{http://www.eso.org/sci/facilities/paranal/telescopes/vlti.html}.

\bibitem{NPOI-url}
\url{http://www2.lowell.edu/npoi/}.

\bibitem{LBT-url}
\url{http://www.lbto.org/}.

\bibitem{GMT-url}
\url{http://www.gmto.org/}.

\bibitem{TMT-url}
\url{http://www.tmt.org/}.

\bibitem{ELT-url}
\url{https://www.eso.org/sci/facilities/eelt/}.

\bibitem{Colossus-2014}
G.~Moretto, J.~R. Kuhn, E.~Thiébaut, M.~Langlois, S.~V. Berdyugina,
  C.~Harlingten, and D.~Halliday, \enquote{New strategies for an extremely
  large telescope dedicated to extremely high contrast: the {C}olossus project,}
  (2014).

\bibitem{pareschi2013dual}
G.~Pareschi, G.~Agnetta, L.~Antonelli, D.~Bastieri, G.~Bellassai, M.~Belluso,
  C.~Bigongiari, S.~Billotta, B.~Biondo, G.~Bonanno \emph{et~al.}, \enquote{The
  dual-mirror small size telescope for the {C}herenkov {T}elescope {A}rray,}
  arXiv preprint arXiv:1307.4962  (2013).

\bibitem{JWST-url}
\url{www.jwst.nasa.gov/}.

\bibitem{DARWIN-url}
\url{http://www.esa.int/Our Activities/Space Science/Darwin overview}.

\bibitem{bolcar-LUVOIRtechno-spie15}
M.~R. Bolcar, K.~Balasubramanian, M.~Clampin, J.~Crooke, L.~Feinberg,
  M.~Postman, M.~Quijada, B.~Rauscher, D.~Redding, N.~Rioux, S.~Shaklan, H.~P.
  Stahl, C.~Stahle, and H.~Thronson, \enquote{Technology development for the
  {A}dvanced {T}echnology {L}arge {A}perture {S}pace {T}elescope ({ATLAST}) as
  a candidate {L}arge {UV}-{O}ptical-{I}nfrared (luvoir) surveyor,}  (2015),
  vol. 9602, pp. 960209--960209--14.

\bibitem{TALC_sauvage}
M.~Sauvage, J.~Amiaux, J.~Austin, M.~Bello, G.~Bianucci, S.~Chesn\'e,
  O.~Citterio, C.~Collette, S.~Correia, G.~A. Durand, S.~Molinari, G.~Pareschi,
  Y.~Penfornis, G.~Sironi, G.~Valsecchi, S.~Verpoort, and U.~Wittrock,
  \enquote{A development roadmap for critical technologies needed for {TALC}: a
  deployable 20m annular space telescope,}  (2016), vol. 9904, pp.
  99041L--99041L--8.

\bibitem{pitman2004remote}
J.~T. Pitman, A.~Duncan, D.~Stubbs, R.~D. Sigler, R.~L. Kendrick, E.~H. Smith,
  J.~E. Mason, G.~Delory, J.~H. Lipps, M.~Manga \emph{et~al.}, \enquote{Remote
  sensing space science enabled by the multiple instrument distributed aperture
  sensor (midas) concept,} in \enquote{Optical Science and Technology, the SPIE
  49th Annual Meeting,}  (International Society for Optics and Photonics,
  2004), pp. 301--310.

\bibitem{Mugnier-p-04}
L.~Mugnier, F.~Cassaing, B.~Sorrente, F.~Baron, M.-T. Velluet, V.~Michau, and
  G.~Rousset, \enquote{{M}ultiple-{A}perture {O}ptical {T}elescopes: some key
  issues for {E}arth observation from a {GEO} orbit,} in \enquote{5th
  International Conference On Space Optics,} , vol. SP-554CNES/ESA (ESA,
  Toulouse, France, 2004), vol. SP-554, pp. 181--187.

\bibitem{Mesrine-p-06}
M.~Mesrine, E.~Thomas, S.~Garin, P.~Blanc, C.~Alis, F.~Cassaing, and
  D.~Laubier, \enquote{High resolution {E}arth observation form {G}eostationary
  orbit by optical aperture synthesys,} in \enquote{Sixth International
  Conference on Space Optics,}  (ESA-CNES, ESA/ESTEC Noordwijk, The
  Netherlands, 2006).

\bibitem{Baron-a-08}
F.~Baron, I.~Mocoeur, F.~Cassaing, and L.~M. Mugnier, \enquote{Unambiguous
  phase retrieval as a cophasing sensor for phased \ array telescopes:
  derivation of an analytical estimator.} J.\ Opt.\ Soc.\ Am.\ A \textbf{25},
  1000--1015 (2008).

\bibitem{Gonsalves-PhaseXretrieval_diversity}
R.~A. Gonsalves, \enquote{Phase retrieval and diversity in adaptive optics,}
  Optical Engineering \textbf{21}, 215829--215829-- (1982).

\bibitem{Mugnier20061}
L.~M. Mugnier, A.~Blanc, and J.~Idier, \enquote{Phase diversity: A technique
  for wave-front sensing and for diffraction-limited imaging,}  (Elsevier,
  2006), pp. 1 -- 76.

\bibitem{Paxman-88}
R.~G. Paxman and J.~R. Fienup, \enquote{Optical misalignment sensing and image
  reconstruction using phase diversity,} J.\ Opt.\ Soc.\ Am.\ A \textbf{5},
  914--923 (1988).
  
\bibitem{Redding-p-98}
D.~C. Redding, S.~A. Basinger, A.~E. Lowman, A.~Kissil, P.~Y. Bely, R.~Burg,
R.~G. Lyon, G.~E. Mosier, M.~Femiano, M.~E. Wilson, R.~G. Schunk, L.~D.
Craig, D.~N. Jacobson, J.~M. Rakoczy, and J.~B. Hadaway, \enquote{Wavefront sensing and control for a {N}ext
	{G}eneration {S}pace {T}elescope,} in \enquote{Space Telescopes and
	Instruments {V},} , vol. 3356 (2) P.~Y. Bely and J.~B. Breckinridge, eds.
(spie, 1998), vol. 3356 (2), pp. 758--772.

\bibitem{Carrara-p-00}
D.~A. Carrara, B.~J. Thelen, and R.~G. Paxman, \enquote{Aberration correction
  of segmented-aperture telescopes by using phase diversity,} in \enquote{Image
  reconstruction from incomplete data,} , vol. 4123 M.~A. Fiddy and R.~P.
  Millane, eds. (spie, 2000), vol. 4123, pp. 56--63.

\bibitem{LLee-p-03}
L.~H. Lee, G.~Vasudevan, and E.~H. Smith, \enquote{Point-by-point approach to
  phase-diverse phase retrieval,} in \enquote{{IR} space telescopes and
  Instruments,} , vol. 4850 J.~C. Mather, ed. (spie, 2003), vol. 4850, pp.
  441--452.

\bibitem{Dean-p-06}
B.~H. Dean, D.~L. Aronstein, J.~S. Smith, R.~Shiri, and D.~S. Acton,
  \enquote{Phase retrieval algorithm for {JWST} {F}light and {T}estbed
  {T}elescope,}  (Proc.\ Soc.\ Photo-Opt.\ Instrum.\ Eng., 2006), vol. 6265,
  pp. 626511--626511--17.

\bibitem{Mocoeur-p-06b}
I.~Moc{\oe}ur, F.~Cassaing, F.~Baron, L.~Mugnier, S.~Hofer, and H.~Thiele,
  \enquote{{Darwin} fringe sensor: experimental results on the {BRISE} bench,}
  in \enquote{Advances in stellar interferometry,} , vol. 6268 J.~D. Monnier,
  M.~Sch\"{o}ller, and W.~C. Danchi, eds. (Proc.\ Soc.\ Photo-Opt.\ Instrum.\
  Eng., 2006), vol. 6268.

\bibitem{Meimon-p-08a}
S.~{Meimon}, E.~{Delavaquerie}, F.~{Cassaing}, T.~{Fusco}, L.~M. {Mugnier}, and
  V.~{Michau}, \enquote{Phasing segmented telescopes with long-exposure phase
  diversity images,} in \enquote{Ground-based and Airborne Telescopes {II},} ,
  vol. 7012 of \emph{Presented at the Society of Photo-Optical Instrumentation
  Engineers (SPIE) Conference} L.~M. Stepp and R.~Gilmozzi, eds. (2008), vol.
  7012 of \emph{Presented at the Society of Photo-Optical Instrumentation
  Engineers (SPIE) Conference}.

\bibitem{JWST_WFSC_2006}
D.~S. Acton, T.~Towell, J.~Schwenker, J.~Swensen, D.~Shields, E.~Sabatke,
  L.~Klingemann, A.~R. Contos, B.~Bauer, K.~Hansen, P.~D. Atcheson, D.~Redding,
  F.~Shi, S.~Basinger, B.~Dean, and L.~Burns, \enquote{Demonstration of the
  {J}ames {W}ebb {S}pace {T}elescope commissioning on the {JWST} testbed
  telescope,}  (2006), vol. 6265, pp. 62650R--62650R--8.

\bibitem{Thurman-JOSAA-11}
S.~T. Thurman, \enquote{Method of obtaining wavefront slope data from
  through-focus point spread function measurements,} J. Opt. Soc. Am. A
  \textbf{28}, 1--7 (2011).

\bibitem{JurlingFienup-JOSAA-14}
A.~S. Jurling and J.~R. Fienup, \enquote{Applications of algorithmic
  differentiation to phase retrieval algorithms,} J. Opt. Soc. Am. A
  \textbf{31}, 1348--1359 (2014).

\bibitem{Carlisle-ApplOpt-15}
R.~E. Carlisle and D.~S. Acton, \enquote{Demonstration of extended capture
  range for {J}ames {W}ebb {S}pace {T}elescope phase retrieval,} Appl. Opt.
  \textbf{54}, 6454--6460 (2015).

\bibitem{JWST_preparing_WFSC}
M.~D. Perrin, D.~S. Acton, C.-P. Lajoie, J.~S. Knight, M.~D. Lallo, M.~Allen,
  W.~Baggett, E.~Barker, T.~Comeau, E.~Coppock, B.~H. Dean, G.~Hartig, W.~L.
  Hayden, M.~Jordan, A.~Jurling, T.~Kulp, J.~Long, M.~W. McElwain, L.~Meza,
  E.~P. Nelan, R.~Soummer, J.~Stansberry, C.~Stark, R.~Telfer, A.~L. Welsh,
  T.~P. Zielinski, and N.~T. Zimmerman, \enquote{Preparing for {JWST} wavefront
  sensing and control operations,}  (2016), vol. 9904, pp. 99040F--99040F--19.

\bibitem{acton2012wavefront}
D.~S. Acton, J.~S. Knight, A.~Contos, S.~Grimaldi, J.~Terry, P.~Lightsey,
  A.~Barto, B.~League, B.~Dean, J.~S. Smith \emph{et~al.}, \enquote{Wavefront
  sensing and controls for the {J}ames {W}ebb {S}pace {T}elescope,}  (Proc.\
  Soc.\ Photo-Opt.\ Instrum.\ Eng., 2012), vol. 8442, p. 84422H.

\bibitem{Noll-ZernikeXpoly}
R.~J. Noll, \enquote{Zernike polynomials and atmospheric turbulence,} J. Opt.
  Soc. Am. \textbf{66}, 207--211 (1976).

\bibitem{Cassaing-p-06b}
F.~Cassaing, B.~Sorrente, L.~Mugnier, G.~Rousset, V.~Michau, I.~Moc{\oe}ur, and
  F.~Baron, \enquote{{BRISE}: a multipurpose bench for cophasing sensors,} in
  \enquote{Advances in stellar interferometry,} , vol. 6268 J.~D. Monnier and
  M.~Sch\"{o}ller, eds. (Proc.\ Soc.\ Photo-Opt.\ Instrum.\ Eng., 2006), vol.
  6268.

\bibitem{Mocoeur:09}
I.~Moc{\oe}ur, L.~M. Mugnier, and F.~Cassaing, \enquote{Analytical solution to
  the phase-diversity problem for real-time wavefront sensing,} Opt. Lett.
  \textbf{34}, 3487--3489 (2009).

\bibitem{vievard2016real}
S.~Vievard, F.~Cassaing, A.~Bonnefois, L.~Mugnier, and J.~Montri,
  \enquote{Real-time alignment and co-phasing of multi-aperture systems using
  phase diversity,} in \enquote{SPIE Astronomical Telescopes+ Instrumentation,}
   (International Society for Optics and Photonics, 2016), pp. 99062Q--99062Q.

\bibitem{goodman2005introduction}
J.~W. Goodman, \emph{Introduction to {F}ourier optics} (Roberts and Company
  Publishers, 2005).

\bibitem{Fontanella-knoxThompson-josaa87}
J.~C. Fontanella and A.~S\`{e}ve, \enquote{Reconstruction of
  turbulence-degraded images using the {K}nox--{T}hompson algorithm,} J. Opt.
  Soc. Am. A \textbf{4}, 438--448 (1987).

\bibitem{Roddier-88}
F.~Roddier, \enquote{Interferometric imaging in optical astronomy,} Physics
  Reports \textbf{170}, 99--166 (1988).

\bibitem{roggemann-imagingThroughTurb-96}
M.~C. Roggemann, B.~M. Welsh, and B.~R. Hunt, \emph{Imaging through turbulence}
  (CRC press, 1996).

\bibitem{knox-imageRecoveryAtm-apj74}
K.~T. {Knox} and B.~J. {Thompson}, \enquote{{Recovery of images from
  atmospherically degraded short-exposure photographs},} Astrophys.\ J.
  \textbf{193}, L45--L48 (1974).

\bibitem{vievard2017developpement}
S.~Vievard, \enquote{D{\'e}veloppement et validation d'un analyseur de surface d'onde
	en plan focal pour un instrument multi-pupille,} Ph.D. thesis, Universit{\'e} Pierre et Marie Curie
(To be published in 2017).

\end{thebibliography}

\providecommand{\inpreparationname}{in preparation}
  \providecommand{\submittedname}{soumis}
  \providecommand{\acceptedname}{accept\'e pour publication}
  \providecommand{\tobepublishedname}{\`a para\^{\i}tre}
  \providecommand{\contractname}{Contrat}
  \providecommand{\conferencedatename}{Conference date: }
  \providecommand{\patent}[2]{Brevet #1 #2}
  \providecommand{\firstabbrevname}{1\textsuperscript{\`ere} }
  \providecommand{\secondabbrevname}{2\textsuperscript{\`eme} }
  \providecommand{\thirdabbrevname}{3\textsuperscript{\`eme} }
  \providecommand{\fourthabbrevname}{4\textsuperscript{\`eme} }
  \providecommand{\fifththabbrevname}{5\textsuperscript{\`eme} }
  \providecommand{\sixththabbrevname}{6\textsuperscript{\`eme} }

\end{document}